\documentclass[nofootinbib,preprintnumbers,prd,superscriptaddress,showpacs,twocolumn]{revtex4}

\usepackage[titletoc]{appendix}
\usepackage{color}
\usepackage{hyperref}

\usepackage[rightcaption]{sidecap}
\usepackage[caption=false]{subfig}
\usepackage{comment}

\usepackage{graphicx}
\usepackage{amsmath}
\usepackage{amsfonts}
\usepackage{amsthm}
\usepackage{mathrsfs}
\usepackage{amssymb}
\usepackage{epsfig}
\usepackage{amscd}
\usepackage{dcolumn}
\usepackage{bm}% bold math
\usepackage{natbib}
\usepackage{url}
\usepackage{xspace}
\usepackage{makecell, array}

\usepackage{dcolumn}

\usepackage{array}
\usepackage{ctable}
\usepackage{multirow}
\usepackage{siunitx}
\usepackage{longtable}
\usepackage{tabularx}
\usepackage{booktabs}

\usepackage{amsthm}
\usepackage{mathrsfs}

\usepackage{epsfig}
\usepackage{amscd}

\usepackage{bm}% bold math
\usepackage{natbib}
\usepackage{url}
\usepackage{xspace}
\usepackage{kantlipsum}
\usepackage{adjustbox}
\usepackage{braket}

\graphicspath{{Graphics/}}

\def\be{\begin{equation}}
\def\ee{\end{equation}}
\def\bea{\begin{eqnarray}}
\def\eea{\end{eqnarray}}

\def\prd{Phys. Rev. D}

\definecolor{vividviolet}{rgb}{0.62, 0.0, 1.0}
\definecolor{amaranth}{rgb}{0.9, 0.17, 0.31}
\definecolor{palatinateblue}{rgb}{0.15, 0.23, 0.89}
\definecolor{brightpink}{rgb}{1.0, 0.0, 0.5}
\definecolor{cornflowerblue}{rgb}{0.39, 0.58, 0.93}
\definecolor{deepcarminepink}{rgb}{0.94, 0.19, 0.22}
\definecolor{radicalred}{rgb}{1.0, 0.21, 0.37}

\hypersetup{ linktoc=all,
    colorlinks, linkcolor={palatinateblue},
    citecolor={brightpink}, urlcolor={amaranth}
}

\begin{document}

\title{Challenging the $\omega_0\omega_a$CDM parametrization through rational expansions in view of DESI data release}

\author{Youri Carloni}
\email{youri.carloni@unicam.it}
\affiliation{Universit\`a di Camerino, Via Madonna delle Carceri, Camerino, 62032, Italy.}
\affiliation{INAF - Osservatorio Astronomico di Brera, Milano, Italy.}
\affiliation{Istituto Nazionale di Fisica Nucleare, Sezione di Perugia, Perugia, 06123, Italy.}

\author{Orlando Luongo}
\email{orlando.luongo@unicam.it}
\affiliation{Universit\`a di Camerino, Via Madonna delle Carceri, Camerino, 62032, Italy.}
\affiliation{INAF - Osservatorio Astronomico di Brera, Milano, Italy.}
\affiliation{Istituto Nazionale di Fisica Nucleare, Sezione di Perugia, Perugia, 06123, Italy.}
\affiliation{SUNY Polytechnic Institute, 13502 Utica, New York, USA.}
\affiliation{Al-Farabi Kazakh National University, Al-Farabi av. 71, 050040 Almaty, Kazakhstan.}

\author{Marek Biesiada}
\email{marek.biesiada@ncbj.gov.pl}
\affiliation{National Centre for Nuclear Research, Pasteura 7, 02-093 Warsaw, Poland.}

\begin{abstract}
In view of the new Dark Energy Spectroscopic Instrument (DESI) 2025 results, we analyze three  types of \emph{Padé cosmology}, based on rational series making use of Padé approximants over the equations of state, namely Padé$^{\omega}$ (0,1) and Padé$^{\omega}$ (1,1), plus a Padé$^{q}$ (0,1), i.e., a rational expansion on the dark energy deceleration parameter, in which where the numerator and denominator orders are incorporated into the above brackets. These scenarios appear alternative dark energy parameterizations with respect to the well-known $\omega_0\omega_a$CDM model, claimed as the most viable model by DESI. Accordingly, we perform Monte Carlo Markov chain (MCMC) analyses with the publicly available \texttt{CLASS} Boltzmann code, including the three Padé cosmology, along with the $\omega_0\omega_a$CDM and $\Lambda$CDM standard pictures. To this end, we combine independent probes from high to low redshifts to obtain reliable constraints on the cosmological parameters of these models and  compare them using statistical selection criteria. \emph{Our results show that Padé cosmology is neither statistically excluded nor worse than the $\omega_0\omega_a$CDM parametrization}. On the contrary, the Akaike Information Criterion (AIC) identifies Padé$^{q}$ (0,1) as \emph{the best-fit model}, with weak evidence against the $\omega_0\omega_a$CDM parameterization, while the Deviance Information Criterion (DIC) provides \emph{strong evidence against the $\omega_0\omega_a$CDM model, favoring Padé (1,1)}. Based on our bounds, we further investigate the evolution of the squared sound speed, revealing that the Padé$^{q}$ (0,1) and Padé$^{\omega}$ (0,1) parameterizations exhibit enhanced stability compared with the other cases here considered and, therefore, describe robust alternatives for the cosmological background.
\end{abstract}


\maketitle
\tableofcontents

%%%%%%%%%%%%%%%%%%%%%%%%%%%%%%%%%%%%%%%%%%%%%%%%%%%%%%%%%%%%%%%%%%%%%%%%%%%%%%%%%%%%%%%%%%%%%%%%

\section{Introduction}

Observations over the past few decades suggest that the Universe is undergoing an accelerated expansion driven by an exotic component, known as dark energy, which constitutes approximately $70\%$ of the total energy budget of the Universe \cite{Padmanabhan:2002ji,Peebles:2002gy,Copeland:2006wr}. The remaining components are dark matter, accounting for about $25\%$, and baryonic matter, which contributes only $\sim 5\%$ \cite{SupernovaCosmologyProject:1997zqe, SupernovaCosmologyProject:1998vns,SupernovaSearchTeam:1998fmf,SupernovaSearchTeam:2003cyd,Huterer:2017buf,Planck:2018vyg}. 

The standard background that accounts for both the energy composition and the observed accelerated expansion is the $\Lambda$CDM paradigm, in which $\Lambda$ represents the cosmological constant, while CDM denotes \emph{cold dark matter}. 

Despite its success in passing a wide range of observational tests, this model has always faced significant challenges, e.g. the cosmological constant problem, the  persistent cosmological tensions \cite{Weinberg:1988cp, Martin:2012bt, SolaPeracaula:2022hpd, Belfiglio:2023rxb,Kamionkowski:2022pkx,Schoneberg:2021qvd,Poulin:2023lkg,Abdalla:2022yfr, DiValentino:2020vvd, Carloni:2025jlk}, etc. that culminated in the recent results,  from the DESI collaboration, favoring the $\omega_0\omega_a$CDM parametrization \cite{DESI:2024mwx, Cortes:2024lgw,Giare:2024gpk,DESI:2025zgx}.

Recent developments, however, showed that this may be only partially true \cite{Colgain:2024xqj,Luongo:2024fww,Carloni:2024zpl,Colgain:2025nzf,Liu:2025myr} since the preferred model may depend on the choice of the combined data sets and/or on the functional structure of dark energy evolution, instead. In addition, it appears still unclear whether the $\omega_0\omega_a$CDM scenario is the most suitable one or not. 

In light of these considerations,  we here perform a further analysis employing \emph{rational expansions of dark energy}, modeled by means of \emph{Padé approximants} as an alternative dark energy equation of state \cite{Gruber:2013wua, Aviles:2014rma, Rezaei:2017yyj, Capozziello:2017ddd,Capozziello:2020ctn,Capozziello:2022jbw}. Precisely, in analogy to the $w$CDM model \cite{Armendariz-Picon:2000nqq}, we employ a Padé (0,1) approach first, then adding one more parameter, we consider a Padé (1,1) scenario, expanding in both the cases the equation of state. Afterwards, motivating our analysis from a genuine thermodynamics, we move on expanding the dark energy deceleration parameter through a Padé (0,1) as a further scheme to analyze, extending the findings found in Ref. \cite{Capozziello:2022jbw}. In all the cases, we thus compare our outcomes with those obtained for the $\omega_0\omega_a$CDM, to check if the so-constructed \emph{Padé cosmology} provides a consistent way to avoid the high redshift divergences that affect Taylor expansions and thereby offer an effective replacement for the $\omega_0\omega_a$CDM scenario. Our Padé cosmology is thus based on alternative parameterizations, namely Padé$^{\omega}$ (0,1), Padé (1,1), and Padé$^{q}$ (0,1), from low to high redshift observations. In particular, we consider the cosmic microwave background (CMB) shift parameters to replace the full CMB data, the second data release (DR2) of baryon acoustic oscillation (BAO) measurements from the DESI mission, the \textit{Pantheon+} Type Ia Supernova (SN Ia), and the observational Hubble data (OHD). By performing a MCMC analysis  using the above cosmic probes, we conclude that Padé approximants provide a viable description of dynamical dark energy, \emph{as they are not statistically excluded}, as certified by exploring AIC and DIC statistical criteria. More precisely, the statistical comparison shows that Padé$^{q}$ (0,1) yields the lowest AIC value, while Padé (1,1) is favored according to  DIC test, albeit \emph{remarkably the $\omega_0\omega_a$CDM parametrization remains disfavored.} Further, based on the results of our analysis, we also compute the squared sound speed and find that Padé$^{q}$ (0,1) and Padé$^{\omega}$ (0,1) exhibit a more stable behavior at late times compared to the other dark energy parameterizations. The overall theoretical and numerical analysis shows that alternative expansions may be used to characterize dark energy, pointing out that the $\omega_0\omega_a$CDM model appears less crucial than what claimed in previous studies \cite{DESI:2024mwx,DESI:2025zgx}. 

The paper is organized as follows. In Sec.~\ref{sec:1}, we introduce the theoretical framework of Padé polynomials used as a parametrization for dark energy. In Sec.~\ref{sec:2}, we outline the experimental setup and detail the cosmic probes employed in our analysis. In Sec.~\ref{sec:3}, we report our findings, comparing the Padé approximants equation of state with the $\omega_0\omega_a$CDM parametrization using the AIC and DIC statistical criteria, and computing the squared sound speed for each model. Finally, in Sec.~\ref{sec:4}, we summarize our results and provide an outlook for future work.

%%%%%%%%%%%%%%%%%%%%%%%%%%%%%%%%%%%%%%%%%%%%%%%%%%%%%%%%%%%%%%%%%%%%%%%%%%%%%%%%%%%%%%%%%%%%%%%%

\section{Dark energy  parameterization}\label{sec:1}

Characterizing dark energy in terms of barotropic fluids can be complicated by the lack of direct information toward the nature of the constituent associated with dark energy itself \cite{Kunz:2012aw, Tsujikawa:2013fta, Wolf:2024eph,Wolf:2025jlc}. On one hand, barotropic fluid is the only one physical description of constituents of the Universe compatible with the homogeneity and isotropy. On the other hand, the detailed equation of state relating fluid's pressure $P$ and density $\rho$ would always be based on a specific physical model of that constituent. 
Hence, approaches to go through this issue includes \emph{a priori} and/or \emph{a posteriori} methods that try to determine the dark energy equation of state, see e.g. \cite{2025PDU....4901965D}. 
One possibility is to use a reasonable assumption that an evolving dark energy equation of state should be an analytical function of the cosmic time \cite{Starobinsky:1998fr,Huterer:1998qv,Sahni:2006pa}. Respecting this constraint, one can in principle use Taylor expansion  in terms of the redshift, scale factor or the Hubble parameter, obtaining usually quite different results, also due to the fact that the expansions are typically limited to very small cosmographic quantities, implying a convergence problem, see e.g. \cite{Cattoen:2007sk,Gruber:2013wua,Bamba:2012cp}. Anyway, this heuristic, phenomenological approach is a promising one as it provides a common point at which observations can meet the theory. Namely, the cosmographic observational tests are able to constrain the coefficients of such expansions, while any potential specific theoretical model of dark energy should be able to predict the values of these coefficients. The best known example of such an approach is the scalar field model of quintessence \cite{RatraPeebles1988,Starobinsky:1998fr}.

In this work, we analyze alternative ways to parametrize dynamical dark energy without relying on specific model properties, instead constructing the background using fundamental series. We then compare these parameterizations with the well-known $\omega_0\omega_a$CDM and $\Lambda$CDM models. Precisely, we proceed as follows: 

\begin{itemize}
    \item[-] We introduce the $\omega_0\omega_a$CDM scenario, which is the most commonly employed parametrization derived from a Taylor expansion in the scale factor, and we recast it in terms of the redshift;
    \item[-] We propose two methods to \emph{stabilize} the expansions at \emph{high redshift} by invoking rational series constructed via Padé reconstruction, which was first introduced in cosmographic analyses in Ref.~\cite{Gruber:2013wua};
    \item[-] We derive a thermodynamically motivated model, as third expansion,  constructed from the simplest viable expansion and expressed through the dark energy deceleration parameter, originally introduced in Ref.~\cite{Capozziello:2022jbw}.
\end{itemize}

These proposals are thus compared among them first and with respect to the $\Lambda$CDM and $\omega_0\omega_a$CDM standard scenarios, by employing the most recent late-time data sets. In so doing, we will demonstrate that the so-constructed \emph{Padé cosmology} is able to better perform than the $\Lambda$CDM and $\omega_0\omega_a$CDM approaches,  opening new avenues toward how dark energy evolves.

\subsection{Standard Taylor expansions: From $z$ expansions to the $\omega_0\omega_a$CDM model}

Any dark energy model can be parameterized by
\begin{equation}
E(z)^{2}= \Omega_m(1+z)^{3} + \Omega_{\text{de}}f(z),
\end{equation}
in a spatially flat Universe. Throughout the paper, we assume spatial flatness of the Universe as well supported by observations \cite{Planck:2018vyg,Efstathiou:2020wem}. Here, $\Omega_m$ and $\Omega_{\text{de}}$ are the present-day dimensionless densities of matter and dark energy, respectively, $E(z)=H(z)/H_0$ is the reduced Hubble parameter, and $f(z)$ describes the redshift evolution of the dark energy component, i.e.,
\begin{equation}
f(z) = \exp\left[\int_{0}^{z} \frac{3(1+\omega(z'))}{1+z'}dz'\right],
\end{equation}
with $\omega(z)$ denoting the equation of state coefficient.

A naive Taylor expansion of $\omega(z)$ coefficient around the present epoch $z=0$
\begin{equation}
    \omega(z) = \sum_{n=0}^{\infty} \omega_n z^{n},
\end{equation}
is known to exhibit divergences at high redshift, thereby restricting its applicability to $z<1$, introducing systematic biases, and reducing numerical accuracy due to slow convergence, see e.g. \cite{Cattoen:2007sk, Aviles:2012ay, Hu:2022udt}.

In fact, truncating the expansion at first order, i.e., $\omega(z)=\omega_0+\omega_1 z$, the corresponding reduced Hubble parameter would read
\begin{equation}
E(z)^{2}=\Omega_{m}(1+z)^{3}+\Omega_{\text{de}}(1+z)^{3(\omega_{0}-\omega_{1}+1)}e^{3\omega_{1}z},\label{eq:TE1}
\end{equation}
explicitly showing a divergence for  high redshifts. 

To solve this issue, a widely-adopted approach is reformulating the expansion in terms of the scale factor, $a=(1+z)^{-1}$ (which is a true physical degree of freedom), as
\begin{equation}
    \omega(a) = \sum_{n=0}^{\infty} \omega_n (1-a)^{n}.
\end{equation}
Then, truncating this series at the first order, yields the well-known $\omega_0\omega_a$CDM  parameterization\footnote{The model was firstly proposed by Chevallier and Polarski \cite{Chevallier:2000qy} and later by Linder \cite{Linder:2002et}, being usually denoted as CPL parametrization. The phantom nature of the equation of state, however, was not predicted by the three authors, suggesting a novel nomenclature under the form of $\omega_0\omega_a$CDM parametrization.},
\begin{equation}\label{defCPL}
    \omega(a) = \omega_0 + \omega_a (1-a),
\end{equation}
which, in light of DESI results obtaining $\omega_0-\omega_a$ contours disjoint from $\Lambda$CDM at 2$\sigma$ level, implies that the $\omega_0\omega_a$CDM parametrization turns out to be taken as a possible viable cosmological model. 

In this case, the corresponding reduced Hubble parameter reads
\begin{equation}
E(a)^{2}=\Omega_{m}a^{-3}+\Omega_{\text{de}}a^{-3 (\omega_{0}+\omega_{a}+1)} e^{-3\omega_a(1-a)},\label{eq:H(a)CPL}
\end{equation}
turning out to be stable at high redshift as 
\begin{equation}
E(z)^{2}=\Omega_{m}(1+z)^{3}+\Omega_{\text{de}}(1+z)^{3 (\omega_{0}+\omega_{a}+1)} e^{-\frac{3\omega_a z}{1+z}}.\label{eq:H(z)CPL}
\end{equation}

\subsection{Going beyond Taylor series: Rational expansions and stabilization of Taylor series}

It appears natural to go beyond the standard Taylor series and to investigate alternative formulations, expressed in function of the redshift, in which the polynomial series is replaced by a \emph{rational polynomial parameterization}, also known as Padé approximant. This prerogative is motivated by the fact that one may invoke the stability of $\omega$ directly on it, without passing through the trick of reformulating everything in function of $a$. 

To do so, we can work out the Padé parameterization or Padé reconstructions, defined on $\omega$ by
\begin{equation}
    \omega(z) = \frac{\sum_{i=0}^{N} a_i z^{i}}{1 + \sum_{j=1}^{M} b_j z^{j}},\label{eq:EOSPADE}
\end{equation}
alleviating the divergence problems of the standard Taylor expansion and extending the redshift domain of convergence.

In particular, for Padé$^{\omega}$ (0,1), the equation of state is given by 
$\omega(z) = \frac{a_0}{1 + b_1 z}$, while for Padé (1,1) is provided by $\omega(z) = \frac{a_0 + a_1 z}{1 + b_1 z}$, 
leading to 
\begin{equation}
    E(z)^{2} =\Omega_m (1+z)^{3} + \Omega_{\text{de}} (1+z)^{\frac{3 (-a_0+b_1-1)}{b_1-1}} (1+b_1 z)^{\frac{3 a_0}{b_1-1}},
\end{equation}
and
\begin{equation}
\begin{split}
    E(z)^{2}=\Omega_m (1+z)^{3}+ \Omega_{\text{de}} (1+z)^{\frac{3 (-a_0+a_1+b_1-1)}{b_1-1}}\\
    \times(1+b_1 z)^{\frac{3 a_1-3 a_0 b_1}{b_1-b_1^2}},
\end{split}
\end{equation}
respectively. Here, for physical consistency, we impose $ a_0 < -1/3$ to reproduce the expected dark energy behavior, and require $b_1 > 0$ to prevent poles in the denominator. Let us note, that formally the $\omega_0\omega_a$CDM model can be recognized as a special case of Padé (1,1) parametrization, in case of small redshifts.

\subsection{Thermodynamic reconstructions from a rational deceleration parameter}

At this stage, we extend our analysis further to explore an alternative use of Padé approximants to characterize the dark energy behavior in a model-independent way, as shown in Ref. \cite{Capozziello:2022jbw}.

Indeed, we can reconstruct the expansion history of the Universe by considering the evolution of kinematic quantities such as the deceleration parameter. We adopt this approach since any cosmological model should remain compatible with the observational constraints arising from the formation of the cosmic structure.

Accordingly, the evolution of cosmic expansion can be described in terms of the deceleration parameter  $q=-1-\frac{\dot{H}}{H^{2}}$, which, up to the epoch of structure formation, assumes the form
\begin{equation}
    q(z)=-1+\frac{3\Omega_m(1+z)^{3}\Omega_{\text{de}}(1+z)\frac{df(z)}{dz}}{2\left(\Omega_m(1+z)^{3}+\Omega_{\text{de}}f(z)\right)}.\label{q}
\end{equation}

Consequently, $q(z)$ is expected to satisfy a set of asymptotic conditions, which are commonly adopted as:
\begin{subequations}
\begin{align}
    q(z) &\rightarrow -1 \; \left( \frac{dq}{dz} > 0 \right) && \text{as } z \rightarrow -1,\label{a} \\
    q(z) &\geq -1 && \forall z,\label{b} \\
    q(z) &\rightarrow \tfrac{1}{2} && \text{for } z \gg 1.\label{c}
\end{align}
\end{subequations}

Here, we consider the dark energy deceleration parameter
\begin{equation}
    q_{\text{de}}(z)=-1+\frac{(1+z)}{H_{\text{de}}}\frac{d H_{\text{de}}}{dz},
\end{equation}
and to describe it we apply the Padé$^{q}$ (0,1) approximant
\begin{equation}
    q_{\text{de}}(z)=\frac{q_{\text{de}}}{1+q_{1}z},
\end{equation}
to have a total $q(z)$ that fulfills all the above requirements.

Therefore, by considering this choice, Eq. \eqref{q} becomes
\begin{equation}
q(z)=\frac{2q_{\text{de}}(1-\Omega_m)(1+z+q_{\text{de}}z)+\Omega_{m}(1+z)^{3}}{2\left[(1-\Omega_m)(1+z+q_{\text{de}}z^{2})+\Omega_m(1+z)^{3}\right]}, 
\end{equation}
where we imposed $q_1=1+q_{\text{de}}$ to respect Eq. \eqref{c}. This expression for the deceleration parameter also respects Eq. \eqref{a}, while Eq. \eqref{b} requires constraining the value of $q_{\text{de}}$ through observational data.

Moreover, this model-independent Padé formulation leads to the following equation of state coefficient
\begin{equation}
        \omega(z)=\frac{1}{3}\left[\frac{2q_{\text{de}}}{1+z(1+q_{\text{de}})}-1\right]
\end{equation}
and to the corresponding reduced Hubble parameter
\begin{equation}
E(z)^{2}=\Omega_m(1+z)^{3}+\Omega_{\text{de}}(1+z+q_{\text{de}}z)^{2}.
\end{equation}

Then, we also examine whether the model-dependent dark energy parameterizations, i.e., Padé${^\omega}$ (0,1), Padé (1,1), and $\omega_0\omega_a$CDM, satisfy the three conditions mentioned above. In particular, Eq.~\eqref{a} is automatically fulfilled for all dark energy parameterizations, whereas we have to check the validity of Eqs.~\eqref{b}-~\eqref{c} after obtaining the MCMC results.

\subsection{The role of sound speed}

The standard theory of structure formation implies that the linear perturbations in Fourier space are given by
\begin{equation}
    \ddot \delta +2H\delta +c_s^2\left(\frac{k^{2}}{a^{2}}-k^{2}_{J}\right)\delta=0,
\end{equation}
where $\delta$ indicates the density perturbations and $k_{J}$ is the physical Jeans wavenumber. 

Here, we can clearly see that the sound speed plays a crucial role in determining the formation of cosmic structures. In particular, comparing the  length scales of perturbations to the Jeans length permits one to characterize at which stage of the Universe's evolution structures may form. This is particularly useful especially in two cases, i) to compare fluids with matter, where the latter clusterizes at all scales, ii) in view of the DESI results that inequivocally showed that dark energy might exhibit a negative sound speed, at least for what concerns the $\omega_0\omega_a$CDM parametrization.  

Nevertheless, considering $\lambda$ as a length and the Jeans length given by
\begin{equation}
    \lambda_J=c_s\sqrt{\frac{\pi}{\rho}},
\end{equation}
ensuring that matter has equation of state $\omega\equiv\frac{P}{\rho}=0$ implies that $\lambda>\lambda_J$ at all scales, simply because $\lambda_J=0$ as consequence of the fact that, for $P = \omega(z)\,\rho$, the squared sound speed is provided by
\begin{equation}
    c^{2}_{s}(z)=\frac{dP}{d\rho}=\omega+\rho\frac{d\omega}{d\rho}.\label{eq:cs2form}
\end{equation}

Consequently, within the framework of the dynamical dark energy parameterizations investigated in this work, we obtain the following analytical expressions for Eq.~\eqref{eq:cs2form}, computed at $z=0$:
\begin{itemize}

  \item[-] Padé$^{\omega}$ (0,1):

  \begin{equation}
    c_s^2(z) = 
    \frac{a_0 (\Omega_m-1) (3 a_0-b_1+3)}{3 (a_0 (\Omega_m-1)-1)};
  \end{equation}

  \item[-] Padé (1,1):

  \begin{equation}
    c_s^2(z) =
  -\frac{(\Omega_m-1) (a_0 (3 a_0-b_1+3)+a_1)}{3 \Omega_m-3 (a_0+1) (\Omega_m-1)};
  \end{equation}

  \item[-] $\omega_0 \omega_a$CDM:

  \begin{equation}
    c_s^2(z) =
    \frac{(\Omega_m-1) (3 \omega_0 (\omega_0+1)+\omega_a)}{3 ((\Omega_m-1) (\omega_0+1)-\Omega_m)};
  \end{equation}

  \item[-] Padé$^{q}$ (0,1):

  \begin{equation}
    c_s^2(z) =-\frac{2 (\Omega_m-1) (1-q_{\rm de}) (q_{\rm de}+1)}{3 \Omega_m (2 q_{\rm de}-1)-6 (q_{\rm de}+1)}.
  \end{equation}

\end{itemize}

%%%%%%%%%%%%%%%%%%%%%%%%%%%%%%%%%%%%%%%%%%%%%%%%%%%%%%%%%%%%%%%%%%%%%%%%%%%%%%%%%%%%%%%%%%%%%%%%

\section{Data used}\label{sec:2}

To constrain the two types of Padé dark energy parameterizations, we employ the following independent cosmological probes: the \textit{Pantheon+} SN Ia, the DESI DR2 measurements, the CMB shift parameters, and the OHD.

The characteristics of these probes are summarized as follows:

\begin{itemize}
\item[-]\textbf{Pantheon+ and SH0ES.}
The \textit{Pantheon+} catalogue comprises $1701$ SNe~Ia over the redshift range $z \in [0,\, 2.3]$, collected from 18 independent subsamples~\citep{Scolnic:2021amr}. This dataset also incorporates the SH0ES Cepheid-calibrated host galaxy distances, which yield an estimate of $H_0$ that is in tension with CMB-based determinations~\cite{Brout:2022vxf}.

When including the SH0ES calibration, the residuals in the SN distance modulus are defined as
\begin{equation}
\label{eq:dmuprime}
\Delta \mu_i =
\begin{cases}
\mu_i - \mu_{{\rm C},i}, & i \in \text{Cepheid hosts}, \\
\mu_i - \mu_{\rm th}(z_i), & \text{otherwise},
\end{cases}
\end{equation}
where $\mu_i$ denotes the measured SN distance modulus, $\mu_{{\rm C},i}$ is the SH0ES Cepheid-based modulus for the host galaxy, and $\mu_{\rm th}(z_i)$ is the theoretical model prediction,
\begin{equation}
\mu_{\rm th}(z_i) = m_i - M = 5\log\!\left[\frac{D_{\rm L}(z_i)}{10\,{\rm pc}}\right],
\end{equation}
with $m_i$ being the rest-frame $B$-band apparent magnitude and $M$ the absolute $B$-band magnitude, treated as a nuisance parameter and marginalized over~\cite{SNLS:2011lii}.
The SN log-likelihood is then
\begin{equation}
\label{eq:likelihood}
\ln \mathcal{L}_{\rm SN} = -\frac{1}{2} \Delta \vec{\mu}^{\,T} C^{-1} \Delta \vec{\mu},
\end{equation}
where the total covariance matrix is $C = C_{\rm SN} + C_{\rm C}$, incorporating both statistical and systematic uncertainties.\footnote{\url{https://github.com/PantheonPlusSH0ES/DataRelease}}

\item[-]\textbf{DESI-BAO DR2.}
For BAO constraints, we utilize the DESI DR2 compilation given in Tab.~\ref{tab:DESIDR2BAO}, consisting of $N_{D}= 19$ measurements: $6$ transverse comoving distances $D_{\rm M}$, $6$ line-of-sight distances $D_{\rm H}$, and $7$ angle-averaged distance $D_{\rm V}$. All quantities are expressed in units of the comoving sound horizon at the baryon drag epoch~\cite{DESI:2025zgx}.

\begin{table*}
\centering
\setlength{\tabcolsep}{2.5em}
\renewcommand{\arraystretch}{1.1}
\begin{tabular}{|l|c|c|c|c|}
\hline\hline
Tracer     & $z_{\rm eff}$ & $D_{\rm M}/r_d$ & $D_{\rm H}/r_d$ & $D_{\rm V}/r_d$ \\
\hline
\hline
BGS & $0.295$ & $-$ & $-$ & $7.942\pm 0.075$  \\
LRG1 & $0.510$ & $13.588\pm 0.167$ & $21.863\pm 0.425$ & $12.720\pm 0.099$  \\
LRG2 & $0.706$ & $17.351\pm 0.177$ & $19.455\pm 0.330$ & $16.050\pm 0.011$  \\
LRG3+ELG1 & $0.934$ & $21.576\pm 0.152$ & $17.641\pm 0.193$ & $19.721\pm 0.091$ \\
ELG2 & $1.321$ & $27.601\pm 0.318$ & $14.176\pm 0.221$ & $24.252\pm 0.174$  \\
QSO & $1.484$ & $30.512\pm 0.760$ & $12.817\pm 0.516$ & $26.055\pm 0.398$  \\
Lya QSO & $2.330$ & $38.988\pm 0.531$ & $8.632\pm 0.101$ & $31.267\pm 0.256$ \\
\hline
\hline
\end{tabular}
\caption{The DESI DR2 dataset provides measurements, along with their corresponding uncertainties, for the bright galaxy survey (BGS), luminous red galaxies (LRG), emission line galaxies (ELG), quasars (QSO), Lyman-$\alpha$ forest quasars (Ly$\alpha$ QSO), and the combined LRG+ELG sample \cite{DESI:2025zgx}.}
\label{tab:DESIDR2BAO}
\end{table*}

The BAO observables are provided by
\begin{subequations}
\begin{align}
\frac{D_{\rm M}(z)}{r_d}&=\frac{c}{r_d}\int ^{z}_{0}\frac{dz'}{H(z')},\\
\frac{D_{\rm H}(z)}{r_d} &= \frac{c}{r_d\, H(z)}, \\
\frac{D_{\rm V}(z)}{r_d} &= \frac{\big[z\,D_{\rm H}(z)\,D_{\rm M}^2(z)\big]^{1/3}}{r_d},
\end{align}
\end{subequations}
where $r_d$ is the comoving size of the sound horizon at baryon drag epoch.

At this point, we can write the BAO log-likelihood as
\begin{equation}
\ln\mathcal{L}_{\rm BAO} \propto -\frac{1}{2} \sum_{i=1}^{N_{D}} \left[ \frac{Y_i - Y(z_i)}{\sigma_{Y_i}} \right]^2,
\end{equation}
where $Y_i$ and $\sigma_{Y_i}$ are the DR2 measurements and uncertainties, and $Y(z_i) \in \{ D_{\rm M}(z_i)/r_d,\, D_{\rm H}(z_i)/r_d,\, D_{\rm V}(z_i)/r_d \}$.
\\
\item[-]\textbf{CMB shift parameters.}
We further include CMB constraints via the shift parameters $\mathcal{R} = 1.7502 \pm 0.0046$ and $l_A = 301.471^{+0.089}_{-0.090}$~\cite{Chen:2018dbv}, where $l_A$ is related to the acoustic angular scale at recombination. The definitions are given as follows
\begin{subequations}
\begin{align}
\mathcal{R}(z_\star) &= \frac{D_{\rm M}(z_\star)\, H_0 \sqrt{\Omega_m}}{c}, \\
l_A(z_\star) &= \pi\,\frac{D_{\rm M}(z_\star)}{r_\star},
\end{align}
\end{subequations}
with the redshift at recombination indicated as $z_\star = 1089.92 \pm 0.25$~\cite{Planck:2018vyg}, and $r_\star$ denoting the comoving sound horizon computed at this epoch.
Then, the CMB log-likelihood takes the form
\begin{equation}
\ln\mathcal{L}_{\rm CMB} \propto -\frac{1}{2} \sum_X \left[ \frac{X - X(z_\star)}{\sigma_X} \right]^2,
\end{equation}
where $X$ refers to $\{\mathcal{R},\, l_A\}$ and $\sigma_X$ denotes the corresponding uncertainties.

\item[-]\textbf{OHD.} Finally, we employ the compilation of OHD, consisting of $N_{O}=32$ measurements of $H(z)$ over the redshift interval $z \in [0.07,\,1.965]$, shown in Tab. \ref{tab:OHD}. Although these measurements exhibit relatively large uncertainties, they are obtained in a fully model-independent manner using the \emph{cosmic chronometer} approach \cite{Jimenez:2001gg}. This method relies on passively evolving red galaxies, for which $H(z)$ is inferred via the differential age technique by evaluating the age difference between pairs of galaxies formed simultaneously but observed at slightly different redshifts, according to
\begin{equation}
    H(z) = -\frac{1}{1+z} \, \frac{\Delta z}{\Delta t}.
\end{equation}

Here, the best-fit parameters are determined by maximizing the following log-likelihood function
\begin{equation}
\label{loglikeOHD}
    \ln \mathcal{L}_{\rm OHD} \propto -\frac{1}{2} \sum_{i=1}^{N_{\rm O}}\left[\dfrac{H_i-H(z_i)}{\sigma_{H_i}}\right]^2 \,,
\end{equation}
where $\sigma_{H_i}$ denotes the uncertainties in the $H(z)$ measurements.

\begin{table}
\centering
\setlength{\tabcolsep}{1.5em}
\renewcommand{\arraystretch}{1.1}
\begin{tabular}{|c|c|c|}
   \hline\hline
    $z$     &$H(z)$ &  References \\
            &[km/s/Mpc]&\\
    \hline\hline
    0.07  & $69.0\pm 19.6$ & \cite{Zhang:2012mp} \\
    0.09    & $69.0 \pm12.0$  & \cite{Jimenez:2001gg} \\
    0.12    & $68.6\pm26.2$  & \cite{Zhang:2012mp} \\
    0.17    & $83.0\pm8.0$   & \cite{Simon:2004tf} \\
    0.179   & $75.0  \pm 4.0$   & \cite{Moresco:2012jh} \\
    0.199   & $75.0\pm5.0$   & \cite{Moresco:2012jh} \\
    0.20    & $72.9\pm29.6$  & \cite{Zhang:2012mp} \\
    0.27    & $77.0\pm14.0$  & \cite{Simon:2004tf} \\
    0.28    & $88.8\pm36.6$  & \cite{Zhang:2012mp} \\
    0.352  & $83.0\pm14.0$  & \cite{Moresco:2016mzx} \\
    0.38  & $83.0\pm13.5$  & \cite{Moresco:2016mzx} \\
    0.4     & $95.0\pm17.0$  & \cite{Simon:2004tf} \\
    0.4004  & $77.0\pm10.2$  & \cite{Moresco:2016mzx} \\
    0.425  & $87.1\pm11.2$  & \cite{Moresco:2016mzx} \\
    0.445  & $92.8 \pm12.9$  & \cite{Moresco:2016mzx} \\
    0.47    & $89.0\pm23.0$     & \cite{Ratsimbazafy:2017vga}\\
    0.4783  & $80.9\pm9.0$   & \cite{Moresco:2016mzx} \\
    0.48    & $97.0\pm62.0$  & \cite{Stern:2009ep} \\
    0.593   & $104.0\pm13.0$  & \cite{Moresco:2012jh} \\
    0.68    & $92.0\pm8.0$   & \cite{Moresco:2012jh} \\
    0.75    & $98.8\pm33.6$     & \cite{Borghi:2021rft}\\
    0.781  & $105.0\pm12.0$  & \cite{Moresco:2012jh} \\
    0.875   & $125.0\pm17.0$  & \cite{Moresco:2012jh} \\
    0.88    & $90.0\pm40.0$  & \cite{Stern:2009ep} \\
    0.9     & $117.0\pm23.0$  & \cite{Simon:2004tf} \\
    1.037   & $154.0\pm20.0$  & \cite{Moresco:2012jh} \\
    1.3     & $168.0\pm17.0$  & \cite{Simon:2004tf} \\
    1.363   & $160.0 \pm 33.6$  & \cite{Moresco:2015cya} \\
    1.43    & $177.0\pm18.0$  & \cite{Simon:2004tf} \\
    1.53    & $140.0\pm14.0$  & \cite{Simon:2004tf} \\
    1.75    & $202.0\pm40.0$  & \cite{Simon:2004tf} \\
    1.965   & $186.5 \pm 50.4$  & \cite{Moresco:2015cya} \\
\hline\hline
\end{tabular}
\caption{OHD catalog with redshifts (first column), measurements with uncertainties (second column), and references (third column).}
\label{tab:OHD}
\end{table}
\end{itemize}

Therefore, by combining all these cosmic probes, the final constraints on the cosmological model parameters are obtained by maximizing the total log-likelihood provided as follows
\begin{equation}
\ln\mathcal{L} = \ln \mathcal{L}_{\rm SN} + \ln \mathcal{L}_{\rm BAO} + \ln \mathcal{L}_{\rm CMB}+\ln \mathcal{L}_{\rm OHD},
\end{equation}
and the uncertainties on the best-fit parameters are assessed through the MCMC analysis, assuming flat priors on the observational data.

%%%%%%%%%%%%%%%%%%%%%%%%%%%%%%%%%%%%%%%%%%%%%%%%%%%%%%%%%%%%%%%%%%%%%%%%%%%%%%%%%%%%%%%%%%%%%%%%

\section{Numerical analysis}\label{sec:3}

In this section, we perform a numerical analysis by applying the cosmic probes introduced in Sec.~\ref{sec:3} to four dynamical dark energy parameterizations, i.e., Padé$^{\omega}$ (0,1), Padé (1,1), $\omega_0\omega_a$CDM, Padé$^{q}$ (0,1) and to the $\Lambda$CDM model.

To do so, we carry out the MCMC analysis with \texttt{CLASS}, a modular Boltzmann solver 
widely used in cosmology for computing CMB anisotropies and related observables. 
Its structure allows straightforward implementation of non-standard physics. 
In our setup, it evolves perturbations in the early Universe, including photons, 
baryons, cold dark matter, neutrinos (with $N_{\rm eff}=3.044$), and our dark energy models.

Then, we adopt information-theoretic criteria to ensure a fair and consistent model comparison. In this way, we verify whether the Padé parameterizations, 
as an alternative phenomenological description of dark energy
are favored over the $\omega_0\omega_a$CDM model, which emerged as a viable, if not preferred
description of late-time cosmology according to the DESI collaboration. 

\subsection{Results and models comparison}

The outcomes of the MCMC analysis are summarized in Tab. \ref{tab:bestfit}, which reports the best-fit values of the cosmological parameters along with their corresponding $1\sigma$ and $2\sigma$ confidence intervals. 

The analysis employs physically motivated priors on the cosmological parameters. Specification of these priors, and the corresponding contour plots, displaying the joint parameter constraints, are presented in Appendix \ref{AppA}.

The behavior of the deceleration parameter is determined from the results obtained by MCMC analysis, with the corresponding plots shown in Appendix \ref{AppB}. Among the considered parameterizations, Padé$^{q}$ (0,1) satisfies all the conditions introduced in Sec. \ref{sec:1}, whereas the others exhibit a different behavior at future times, i.e., for $z \rightarrow -1$.

\begin{figure}[t]
    \centering
        \includegraphics[width=0.48\textwidth]{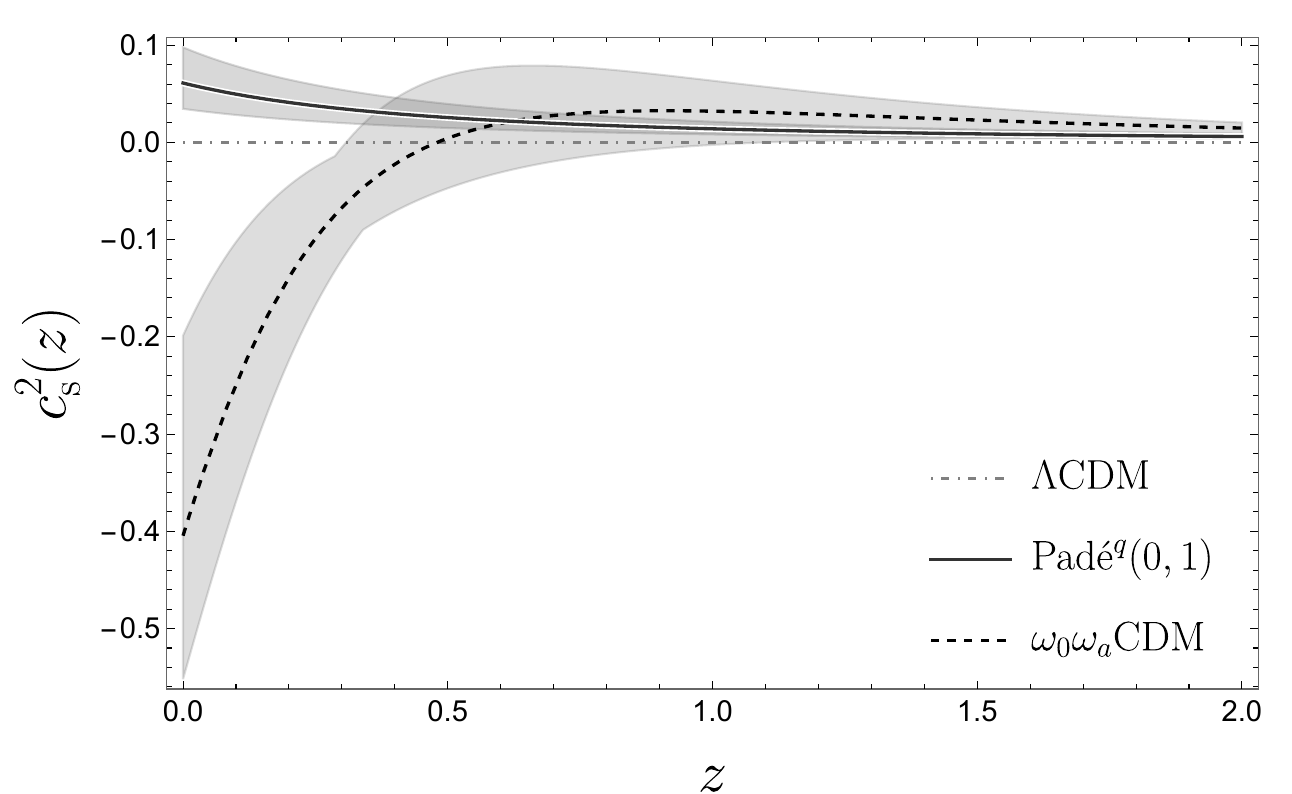}
        \includegraphics[width=0.48\textwidth]{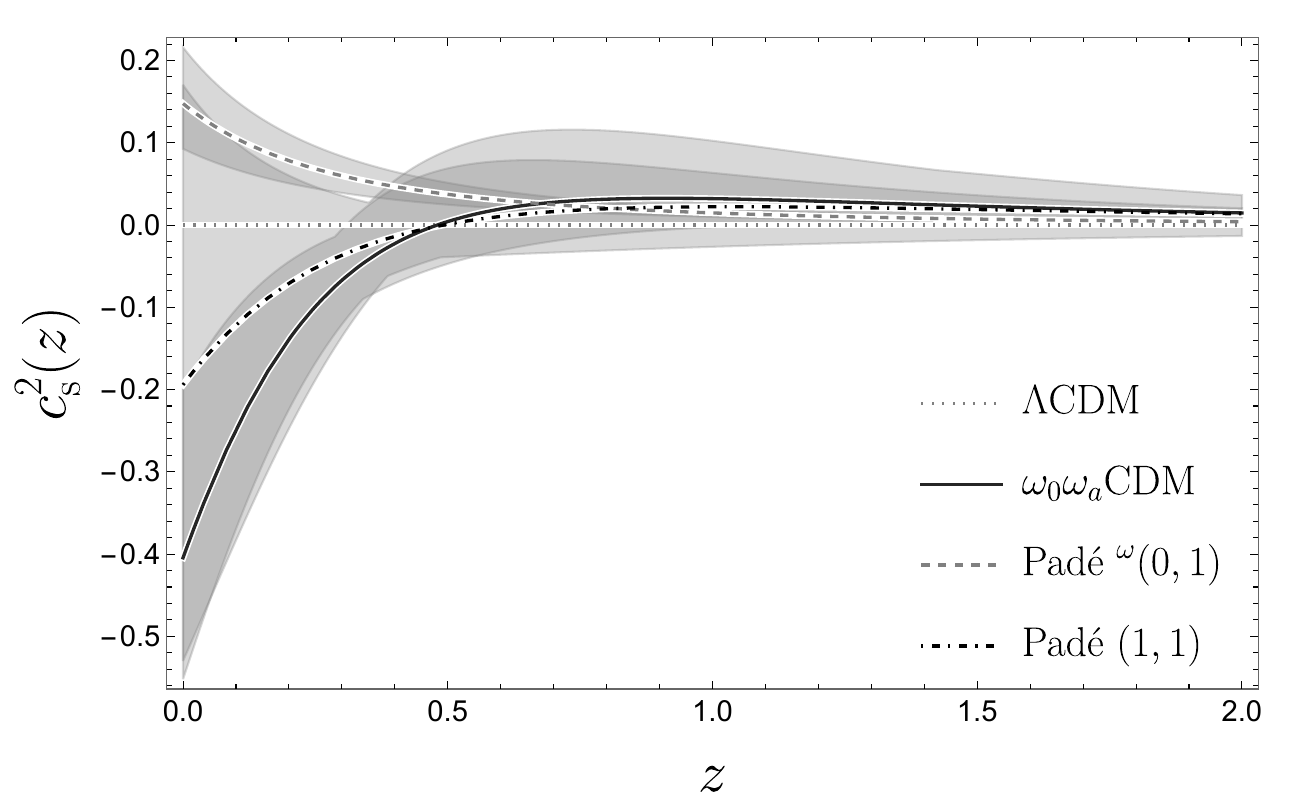}

    \caption{Sound speed as a function of redshift, obtained from the MCMC analysis of the proposed Padé dark energy parameterization, compared with the $\Lambda$CDM and $\omega_0\omega_a$CDM models. The shaded regions denote the 1$\sigma$ confidence intervals.}
    \label{fig:cs2}
\end{figure}

In addition, the squared sound speed, represented in Fig. \ref{fig:cs2}, shows that $\omega_0\omega_a$CDM and Padé (1,1) lead to negative values at late times, indicating unphysical behavior. In contrast, such behavior is avoided in the Padé$^{q}$ (0,1) and Padé$^{\omega}$ (0,1) frameworks, which remain stable also at late times.
 
\begin{table*}[ht]
\centering
\setlength{\tabcolsep}{0.4em}
\renewcommand{\arraystretch}{1.4}
\begin{tabular}{lccccc}
\hline\hline
Parameter & Padé$^{\omega}$ (0,1) & Padé (1,1) & $\omega_0\omega_a$CDM & Padé$^{q}$ (0,1) & $\Lambda$CDM  \\
\hline\hline
$H_0~({\rm km/s/Mpc})$ & $70.80^{+0.82(1.41)}_{-0.69(1.52)}$ & $70.24^{+0.73(1.29)}_{-0.69(1.36)}$ & $70.22^{+0.60(1.39)}_{-0.66(1.39)}$ & $70.89^{+0.69(+1.38)}_{-0.69(-1.38)}$ & $69.09^{+0.35(+0.68)}_{-0.43(-0.67)}$\\
$\Omega_m$ & $0.289^{+0.004(0.008)}_{-0.004(0.008)}$ & $0.295^{+0.004(0.008)}_{-0.004(0.008)}$ & $0.295^{+0.004(0.008)}_{-0.004(0.008)}$ & $0.289^{+0.003(0.007)}_{-0.003(0.007)}$ & $0.294^{+0.003(0.006)}_{-0.003(0.006)}$ \\
$\Omega_{\text{de}}$ & $0.710^{+0.004(0.008)}_{-0.004(0.008)}$ & $0.705^{+0.005(0.008)}_{-0.004(0.009)}$ & $0.705^{+0.004(0.008)}_{-0.004(0.008)}$ & $0.710^{+0.003(0.007)}_{-0.004(0.007)}$ & $0.705^{+0.003(0.006)}_{-0.003(0.006)}$ \\
$a_0$ & $-1.05^{+0.016(0.034)}_{-0.017(0.033)}$ & $-0.984^{+0.039(0.075)}_{-0.039(0.075)}$ & $-$ & $-$ & $-$ \\
$b_1$ & ${3.7\times 10^{-4}}^{+0.00009(0.0002)}_{-0.0001(0.0002)}$ & $0.311^{+0.104(0.330)}_{-0.203(0.275)}$ & $-$ & $-$ & $-$\\
$a_1$ & $-$ & $-0.512^{+0.215(0.378)}_{-0.174(0.388)}$ & $-$ & $-$ & $-$ \\
$\omega_0$ & $-$ & $-$ & $-0.950^{+0.039(0.092)}_{-0.044(0.093)}$ & $-$ & $-$\\
$\omega_a$ & $-$ & $-$ & $-0.426^{+0.186(0.396)}_{-0.158(0.401)}$ & $-$ & $-$\\
$q_{\text{de}}$ & $-1.07^{+0.026(0.051)}_{-0.024(0.046)}$ & $-0.975^{+0.057(0.108)}_{-0.057(0.110)}$ & $-0.929^{+0.075(0.146)}_{-0.074(0.152)}$ & $-1.05^{+0.019(0.043)}_{-0.022(0.039)}$ & $-1$\\
$\ln\mathcal L_{m}$ & $-673.43$ & $-670.82$ & $-671.44$ & $-670.63$ & $-676.78$ \\
\hline\hline
\end{tabular}
\caption{Best-fit parameters with $1\sigma$ ($2\sigma$) uncertainties and total log-likelihood values of Padé$^{\omega}$ (0,1), Padé (1,1), $\omega_0\omega_a$CDM, Padé$^{q}$ (0,1), and $\Lambda$CMD models, obtained by fitting SNe Ia, BAO, CMB, and OHD datasets.}
\label{tab:bestfit}
\end{table*}

As mentioned above, we performed a quantitative comparison of DESI supported $\omega_0\omega_a$CDM parametrization with the Padé approximants. 

For this purpose, we apply a set of well-established model selection criteria. In cosmology, the most widely used diagnostics are the AIC and DIC criteria \cite{Kunz:2006mc,Liddle:2007fy,Biesiada:2007um,Szydlowski:2005kv,Szydlowski:2006pz}. These quantities, extensively used in regression model assessment \cite{Burnham:2002book}, are defined as
\begin{subequations}
\label{eq:AIC_DIC}
\begin{align}
& \text{AIC} \equiv -2\ln \mathcal{L}_{m} + 2d, \\
& \text{DIC} \equiv -2\ln\mathcal{L}_{m} + 2p,
\end{align}
\end{subequations}
where $\ln \mathcal{L}_{m}$ denotes the maximum log-likelihood, see Tab. \ref{tab:bestfit}, $d$ is the number of free parameters in a given model, and $p=\braket{-2\ln\mathcal{L}}+2\ln\mathcal{L}_{m}$, where brackets denote the average over the posterior distribution of parameters. The quantity $p$, called Bayesian complexity, has the meaning of an effective number of parameters of a given model constrained by the data~\cite{Liddle:2007fy}. One can note formal similarity of mathematical expressions for AIC and DIC.

The conceptual foundation of these estimators relies on two probability distributions: $h(x)$, the true but unknown distribution, and $g(x|\theta)$, a parametric model characterized by the parameter set $\theta$. An optimal parameter set $\theta_{\min}$ exists that minimizes the discrepancy between $g(x|\theta)$ and $h(x)$~\cite{Sugiura01011978}. The AIC is derived from an approximate minimization of the Kullback–Leibler information entropy, which is one of the ways to quantify the discrepancy between the true distribution of the data and the model distribution.  By using the effective number of parameters, the DIC overcomes the problem of the AIC that it does not discount parameters which are unconstrained by the data. In other words, in contrast to AIC, DIC penalizes only those parameters that are effectively constrained by the data~\cite{Liddle:2007fy}.

Since $h(x)$ is unknown, the absolute values of AIC, or DIC for a single model are not directly informative. Instead, model comparison is performed through relative differences,
\begin{equation}
\Delta X = X_i - X_0, \quad X_i \in {\text{AIC, DIC}},
\end{equation}
where $X_0$ denotes the minimum value among the competing models. In our analysis, we therefore adopt as the baseline the model that minimizes the information criterion, i.e., the one for which $\Delta X=0$. 

The magnitude of $\Delta X$ provides a quantitative measure of the strength of preference: $\Delta X \geq 2$ (weak), $\Delta X \geq 6$ (moderate), and $\Delta X \geq 10$ (strong) \cite{CosmoVerseNetwork:2025alb}.

\begin{table*}[ht]
\centering
\setlength{\tabcolsep}{0.5em}
\renewcommand{\arraystretch}{1.8}
\begin{tabular}{llllll}
\hline\hline
$X (\Delta X)$ & Padé$^{\omega}$ (0,1) & Padé (1,1) & $\omega_0\omega_a$CDM & Padé$^{q}$ (0,1) & $\Lambda$CDM\\
\hline\hline
AIC ($\Delta$AIC) & $1351(8)$ & $1348(5)$ & $1347(4)$ & $1343(0)$ & $1356(13)$ \\

DIC ($\Delta$DIC) & $1357(5)$ & $1352(0)$ & $1383(31)$ & $1373(21)$ & $1438(86)$\\
\hline\hline
\end{tabular}%
%}
\caption{Model selection criteria and relative differences with respect to the baseline for the considered parametrizations.}
\label{tab:stat}
\end{table*}

As shown in Tab.~\ref{tab:stat}, the Padé$^{q}$ (0,1) parameterization emerges as the best-fit model according to the AIC, while the $\omega_0\omega_a$CDM model is weakly disfavored. In contrast, the DIC criterion favors the Padé (1,1) model, with a weak preference over Padé$^{\omega}$ (0,1) and a strong preference over the other parameterizations, including, once again, the $\omega_0\omega_a$CDM model. The $\Lambda$CDM paradigm, on the other hand, remains strongly disfavored, in agreement with the findings reported by the DESI collaboration.

These results suggest that the Padé approximant method provides a viable alternative to the $\omega_0\omega_a$CDM parameterization for describing the behavior of dark energy. In particular, the Padé$^{\omega}$ (0,1) and Padé$^{q}$ (0,1) parameterizations exhibit a more stable behavior compared to the other dynamical dark energy models, showing a positive squared sound speed at late times.

%%%%%%%%%%%%%%%%%%%%%%%%%%%%%%%%%%%%%%%%%%%%%%%%%%%%%%%%%%%%%%%%%%%%%%%%%%%%%%%%%%%%%%%%%%%%%%%%

\section{Final outlooks}\label{sec:4}

The DESI mission has shown that dark energy seems to exhibit a unexpected evolution in terms of redshifts, indicating that parameterizing dark energy in terms of the redshift or scale factor may reproduce well data even at low redshifts. This fact was a \emph{unicum} in the recent developments in cosmology, where experimental tests always preferred the non-evolving cosmological constant and the corresponding $\Lambda$CDM paradigm, especially while combining low with high cosmic probes. 

Since several exceptions and criticisms against DESI have been raised immediately the DESI release, it appears natural to wonder whether dark energy, under the form of a parametrization, is or is not the final framework to describe dark energy at all stages of Universe evolution. 

In this respect, as Taylor series in $z$ or $a$ simply suggest that the dark energy equation of state is analytical, it appears natural to work with rational approximations that, instead, are constructed to be more stable as the redshift or scale factor increase. 

The main purpose of this work was, in fact, to test rational approximations, constructed starting from the well-known Taylor series, namely coincident with these expansions, up to a given order and by means of the well-established technique offered by Padé polynomials. More precisely, we here adopted Padé$^{\omega}$ (0,1),  (1,1), and Padé$^{q}$ (0,1) parameterizations of dark energy as alternatives to the $\omega_0\omega_a$CDM model, using DESI DR2 data, i.e., the second DESI release, recently released. Padé polynomials provide a consistent way to regulate the high-redshift divergence of Taylor expansions expressed in redshift, whereas the $\omega_0\omega_a$CDM model corresponds to a Taylor expansion around the scale factor normalized to unity today. In other words, we employed some sort of \emph{Padé cosmology} at background, invoking expansions on the equations of state and the dark energy deceleration parameter.

Quite remarkably, \emph{our analysis showed that $\omega_0\omega_a$CDM is not significantly supported by the data, even though the latter is statistically more viable than the $\Lambda$CDM paradigm}.
Surprisingly, the $\omega_0\omega_a$CDM parametrization does not emerge as the best-fit model. According to the AIC, it is mildly disfavored with respect to the Padé$^{q}(0,1)$ parametrization. Moreover, the DIC provides strong evidence against $\omega_0\omega_a$CDM, suggesting the Padé (1,1) parametrization as the preferred model. This indicates that the DESI preference for the $\omega_0\omega_a$CDM is not so robust as naively claimed by the collaboration and recent findings \cite{Reboucas:2024smm,Giare:2024gpk,DESI:2024mwx,DESI:2025zgx}, in line of recent criticisms that showed that there is no clear indication toward the model \cite{Luongo:2024fww,Carloni:2024zpl,Colgain:2024xqj,Colgain:2025fct,Colgain:2025nzf}, instead. However, our outputs certified that the $\Lambda$CDM model can be either preferred or discarded in favor of more complicated models than the $\omega_0\omega_a$CDM parametrization. 

Nevertheless, as already mentioned, the $\omega_0\omega_a$CDM parametrization is equivalent to a special case of Padé (1,1) with $a_0 = \omega_0; \, a_1 = \omega_0 + \omega_a; \, b_1 =1$. Then, our results show that DESI support for $\omega_0\omega_a$CDM can be in fact a support for Padé (1,1), but with a constrained range of coefficients. 

Thus, the values of the best fitted cosmological parameters $H_0$, $\Omega_m$, $\Omega_{de}$ are almost identical for these two parametrizations of cosmic equation of state. Preference of Padé (1,1) according to DIC supports this assertion. Namely, DIC uses the notion of effective number of parameters supported by the data and hence gives preference to full Padé (1,1) case over its constrained case i.e. $\omega_0\omega_a$CDM.  

To show this, we employed a MCMC simulation based on the free available CLASS code, modifying the background, and adopting SH0ES data points, together with the OHD catalog of data points and CMB shift parameters. Clearly, every analysis has been carried out, including the DESI DR2 points, consisting of eight points with refined errors, with respect to the DR1 release. The covariance matrix has not been used, following the usual recipe that does not take it into account, in analogy to the original DESI paper, while uncertainties have been computed up to $2\sigma$ confidence levels. 

Then, the best fitted models have been compared using AIC and DIC information criteria.
The overall analysis, as claimed above, suggested that i) \emph{the best benchmark paradigm is not a simple parametrization in terms of $a$}, as claimed by DESI, ii) \emph{the negative sound speed is not the favored scenario that seems to be compatible with DESI data}, if we consider the AIC criterion for the comparison.

Future works, combining DESI DR2 with other independent cosmic probes, will be essential to further test Padé parameterizations and to clarify whether $\omega_0\omega_a$CDM can still be regarded as the best dark energy model, as claimed by the DESI collaboration. The most important output from our work was the need of extending the standard parameterizations, claimed as definitive by the collaboration and previous literature. However, understanding the nature of negative sound speed for dark energy in both structure formation and cosmic evolution would shed light on the nature of dark energy. 

Remarkably, dark energy was supposed to be nearly constant, with an identically vanishing sound speed. Departing from this scenario toward models exhibiting phantom-like behavior demands careful and rigorous physical justification.

%%%%%%%%%%%%%%%%%%%%%%%%%%%%%%%%%%%%%%%%%%%%%%%%%%%%%%%%%%%%%%%%%%%%%%%%%%%%%%%%%%%%%%%%%%%%%%%%

\section*{Acknowledgements}
YC and OL acknowledge Roberto Della Ceca and the National Institute for Astrophysics (INAF) for partial financial support. OL acknowledges financial support from the Fondazione  ICSC, Spoke 3 Astrophysics and Cosmos Observations. National Recovery and Resilience Plan (Piano Nazionale di Ripresa e Resilienza, PNRR) Project ID CN$\_$00000013 "Italian Research Center on  High-Performance Computing, Big Data and Quantum Computing"  funded by MUR Missione 4 Componente 2 Investimento 1.4: Potenziamento strutture di ricerca e creazione di "campioni nazionali di R$\&$S (M4C2-19 )" - Next Generation EU (NGEU)
GRAB-IT Project, PNRR Cascade Funding
Call, Spoke 3, INAF Italian National Institute for Astrophysics, Project code CN00000013, Project Code (CUP): C53C22000350006, cost center STI442016.

%%%%%%%%%%%%%%%%%%%%%%%%%%%%%%%%%%%%%%%%%%%%%%%%%%%%%%%%%%%%%%%%%%%%%%%%%%%%%%%%%%%%%%%%%%%%%%%%

%%%%%%%%%%%%%%%%%%%%%%%%%%%%%%%%%%%%%%%%%%%%%%%%%%%%%%%%%%%%%%%%%%%%%%%%%%%%%%%%%%%%%%%%%%%%%%%%

\appendix

\section{Contour plots}\label{AppA}

\begin{figure*}[htb!]
    \centering
    \includegraphics[width=0.5\textwidth]{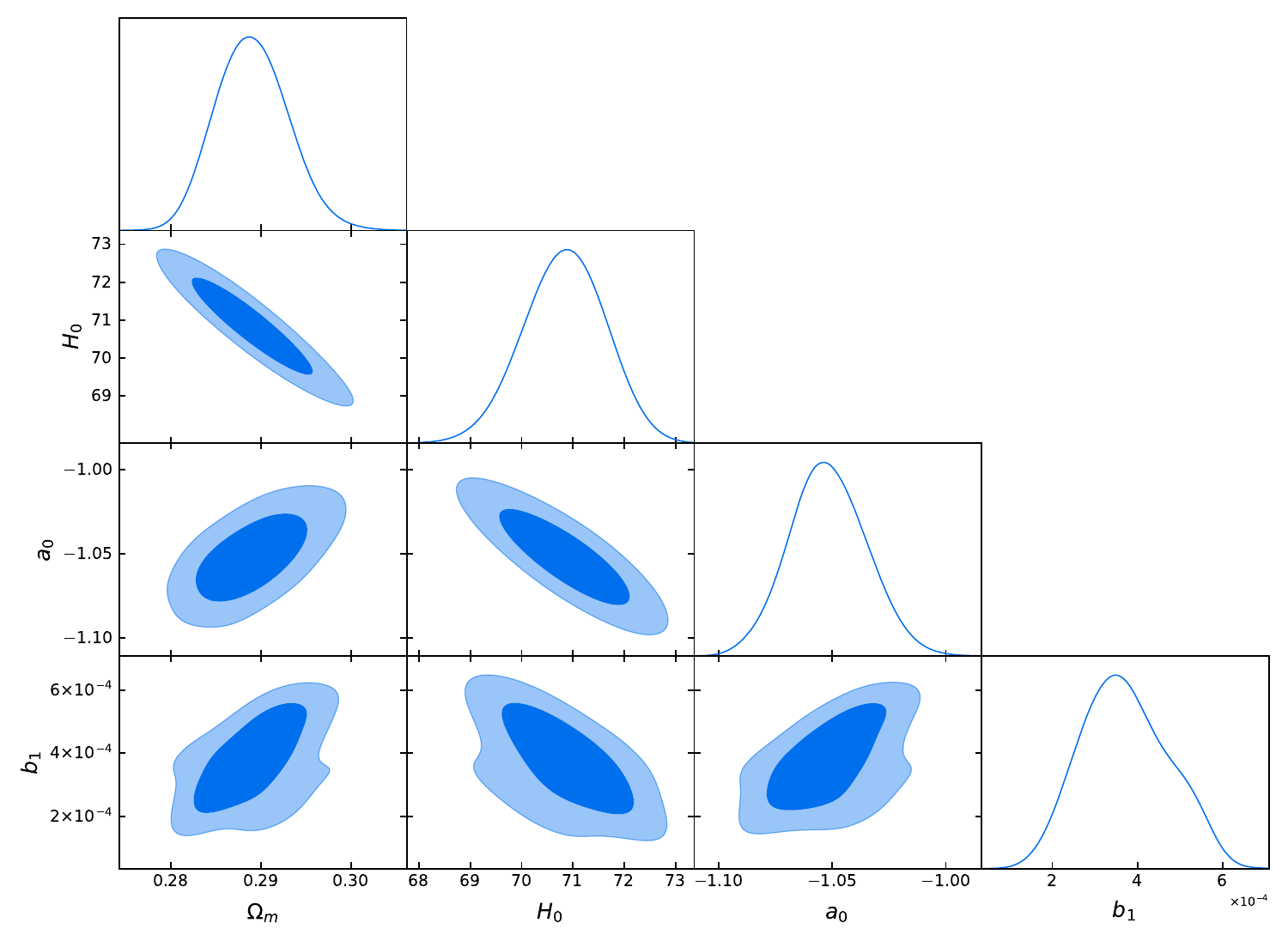}\hfill
    \includegraphics[width=0.5\textwidth]{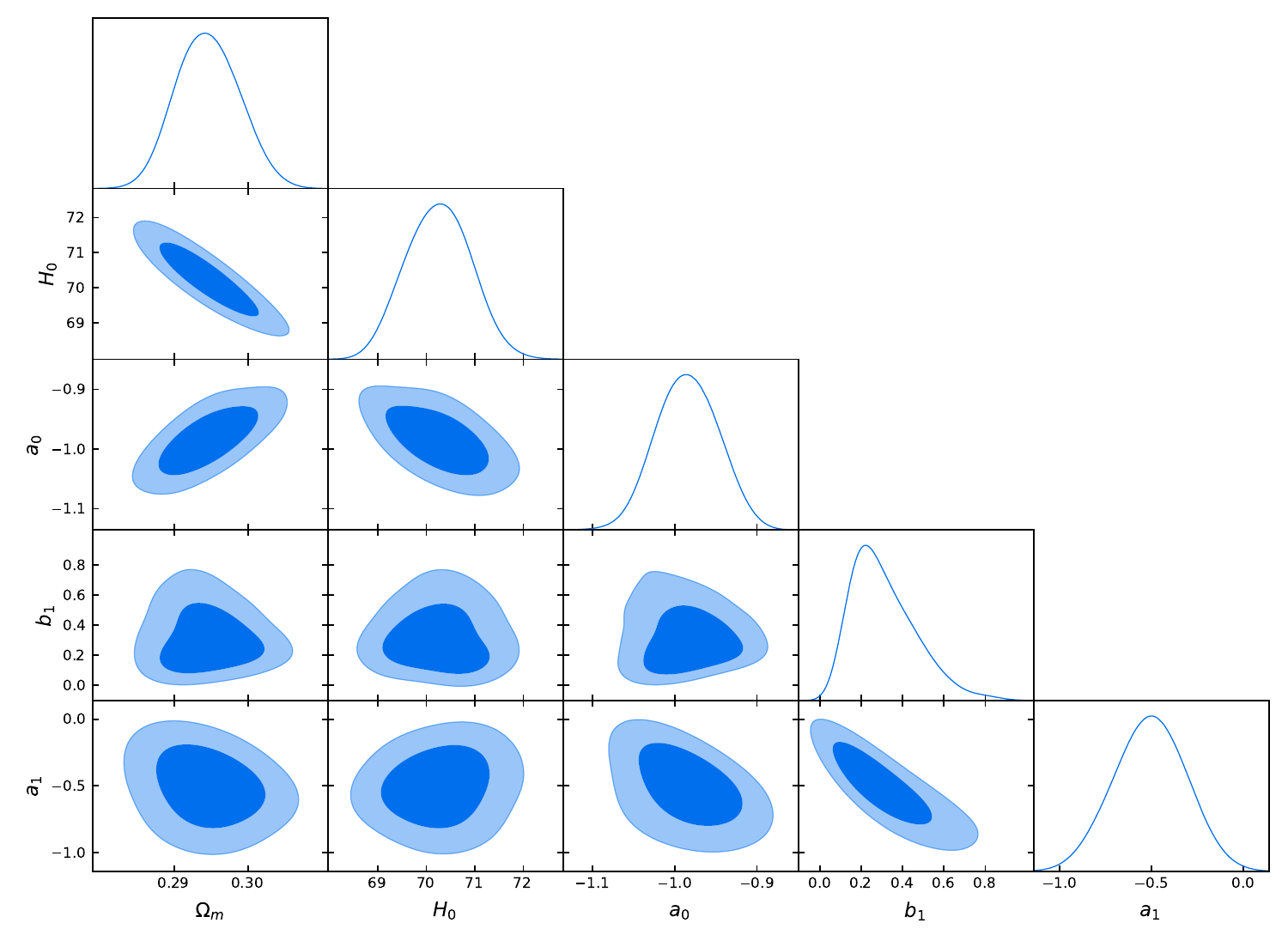}\\[1ex]
    \includegraphics[width=0.5\textwidth]{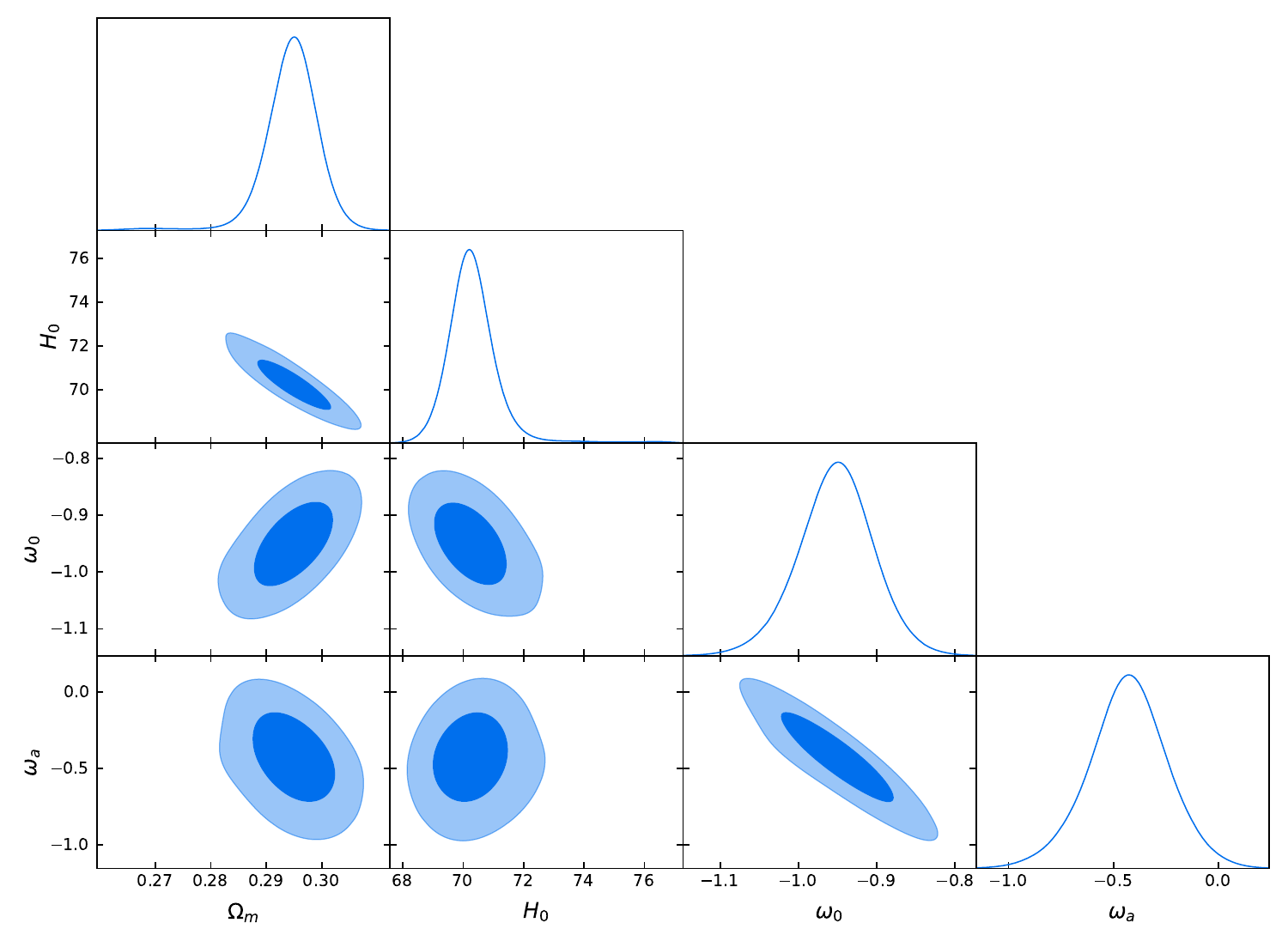}\hfill
     \includegraphics[width=0.5\textwidth]{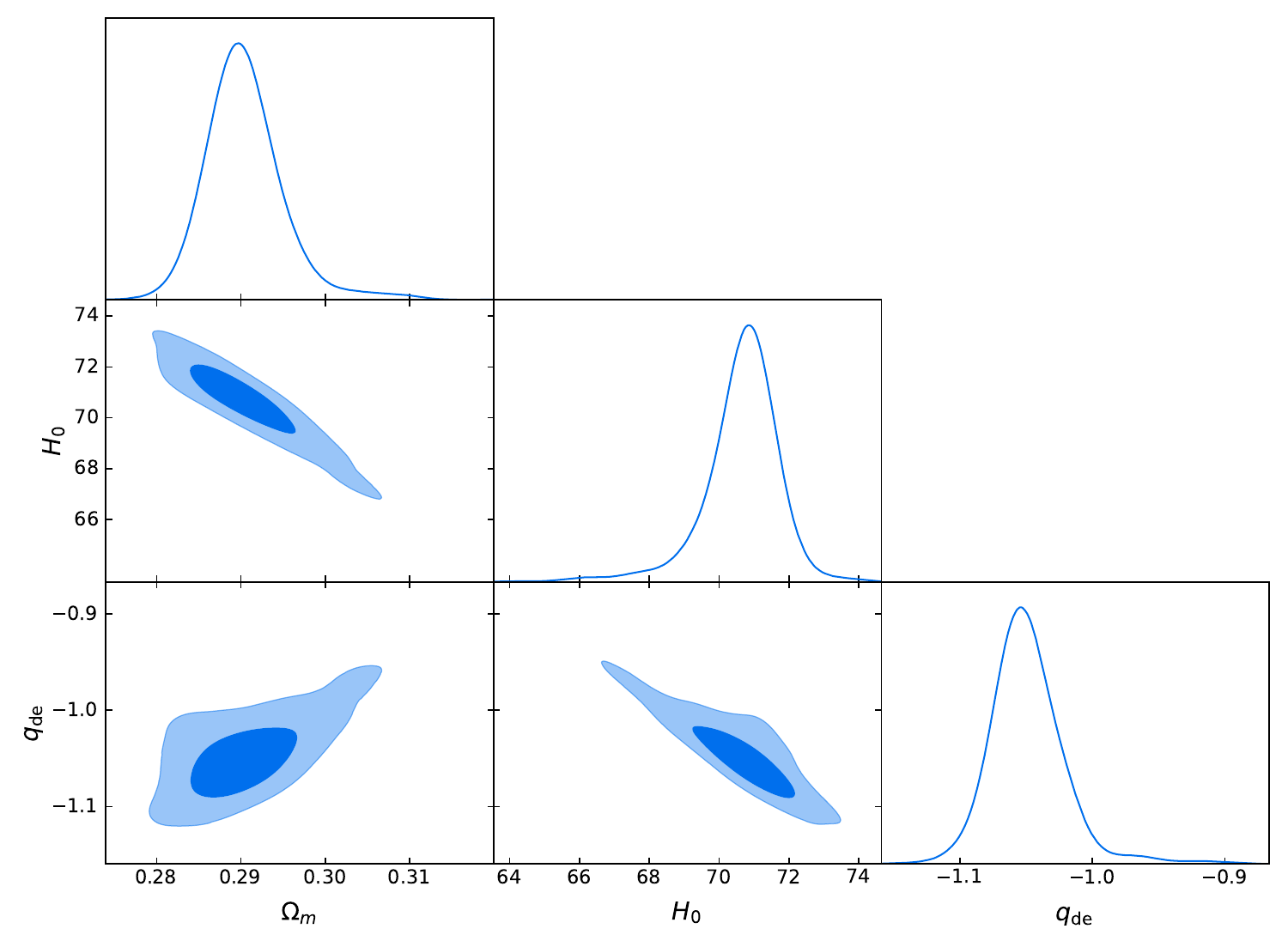}
    \caption{Contour plots of the cosmological parameters for the three dark energy parameterizations considered in this work, obtained with \texttt{GetDist}. 
    The top row corresponds to Padé$^{\omega}$ (0,1) and Padé (1,1), while the bottom panel shows $\omega_0\omega_a$CDM and Padé$^{q}$ (0,1).}
    \label{fig:contours}
\end{figure*}

In this appendix, we report the contour plots obtained with \texttt{GetDist}, see Fig. \ref{fig:contours}, together with the details of the MCMC analyses carried out for the different parameterizations. For each model, we specify the adopted priors.

\begin{itemize}
    \item[-] Padé$^{\omega}$ (0,1): 
    \begin{equation}
    \begin{split}
    &\Omega_{m}\in [0.1,0.9],\, H_{0}\in [20,100]\, \text{km/s/Mpc},\\
    &a_0\in [-3,-0.5],\,b_1\in [0,0.01].
    \end{split}
    \end{equation}

    \item[-] Padé (1,1):
    
    \begin{equation}
    \begin{split}
    &\Omega_{m}\in [0.1,0.9],\, H_{0}\in [20,100]\, \text{km/s/Mpc},\\
    &a_0\in [-3,-0.5],\, b_1\in [0,0.3],\, b_2\in [-1,0].
    \end{split}
    \end{equation}
    
    \item[-] $\omega_0\omega_a$CDM:
    
    \begin{equation}
    \begin{split}
    &\Omega_{m}\in[0.1,0.9],\, H_{0}\in [20,100] \, \text{km/s/Mpc},\\
    &\omega_0\in [-3,1],\, \omega_a\in [-3,0.75].
    \end{split}
    \end{equation}      \\
    
    \item[-] Padé$^{q}$ (0,1):
    
    \begin{equation}
    \begin{split}
     &\Omega_{m}\in[0.1,0.9],\, H_{0}\in [20,100] \, \text{km/s/Mpc},\\
     &q_{\text{de}}\in [-1.2,-0.8].
    \end{split}
    \end{equation}
\end{itemize}

\section{Deceleration parameter evolution}\label{AppB}

\begin{figure*}[t]
    \centering
    \includegraphics[width=0.5\textwidth]{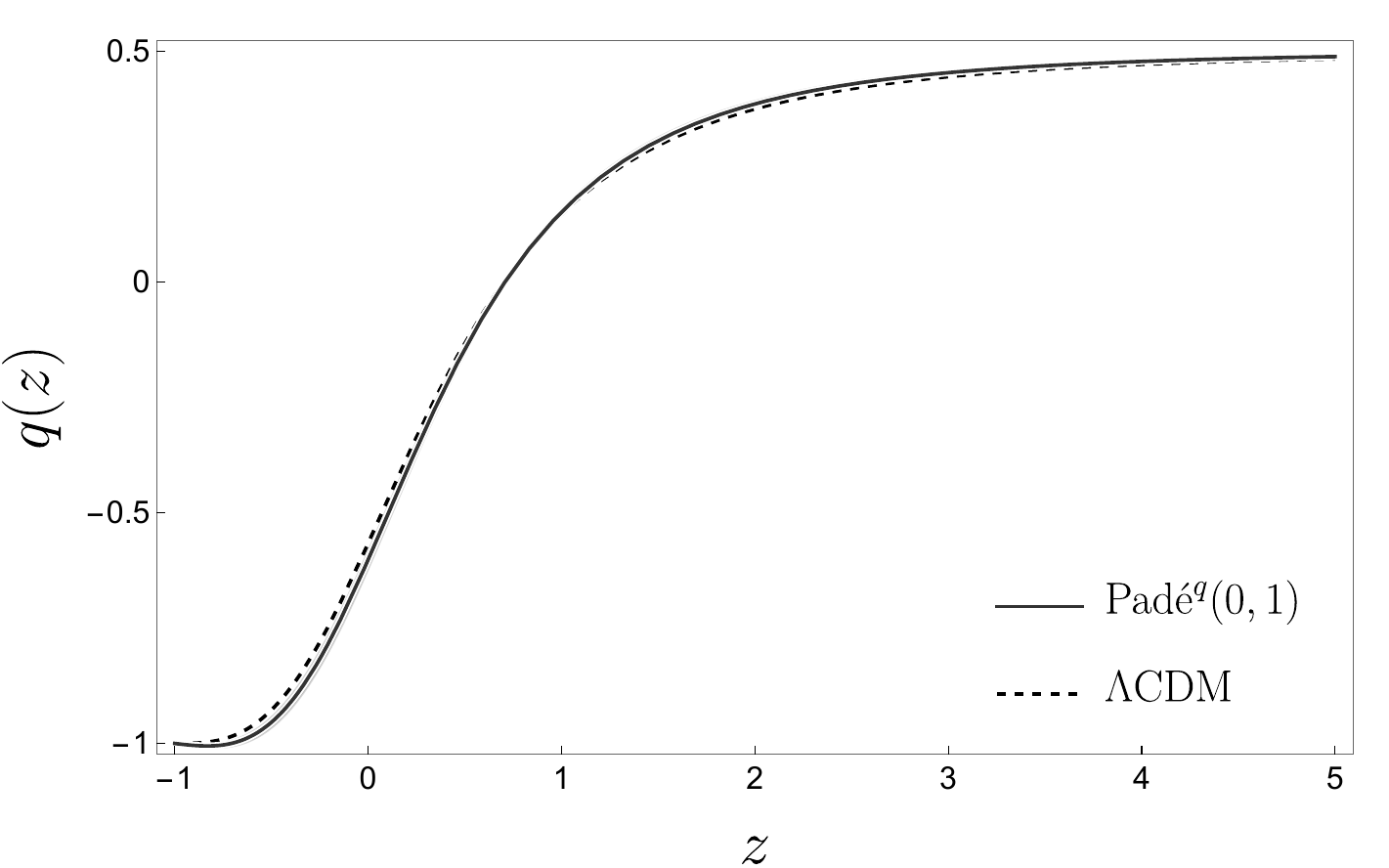}\hfill
    \includegraphics[width=0.5\textwidth]{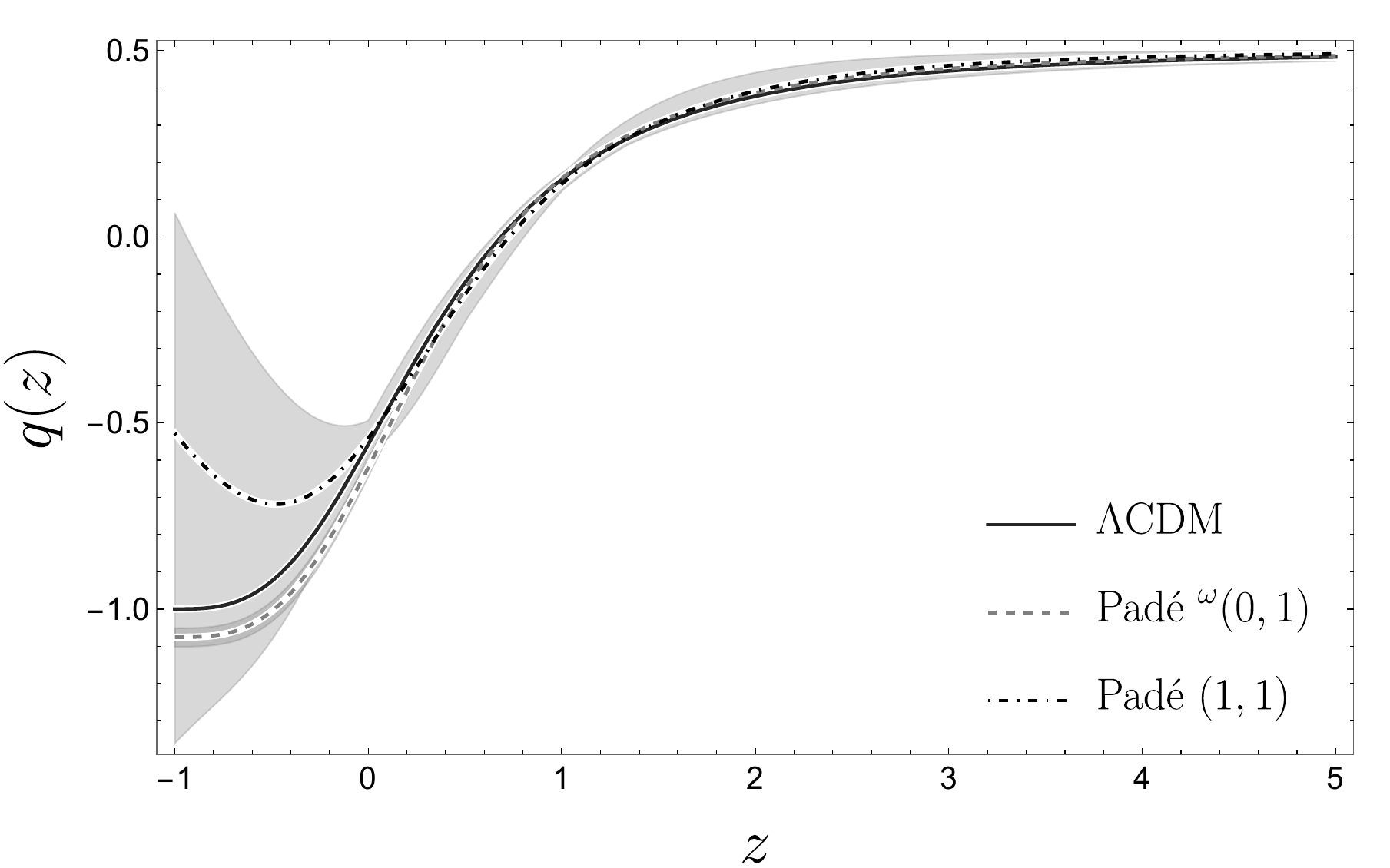}\\[1ex]
    \includegraphics[width=0.5\textwidth]{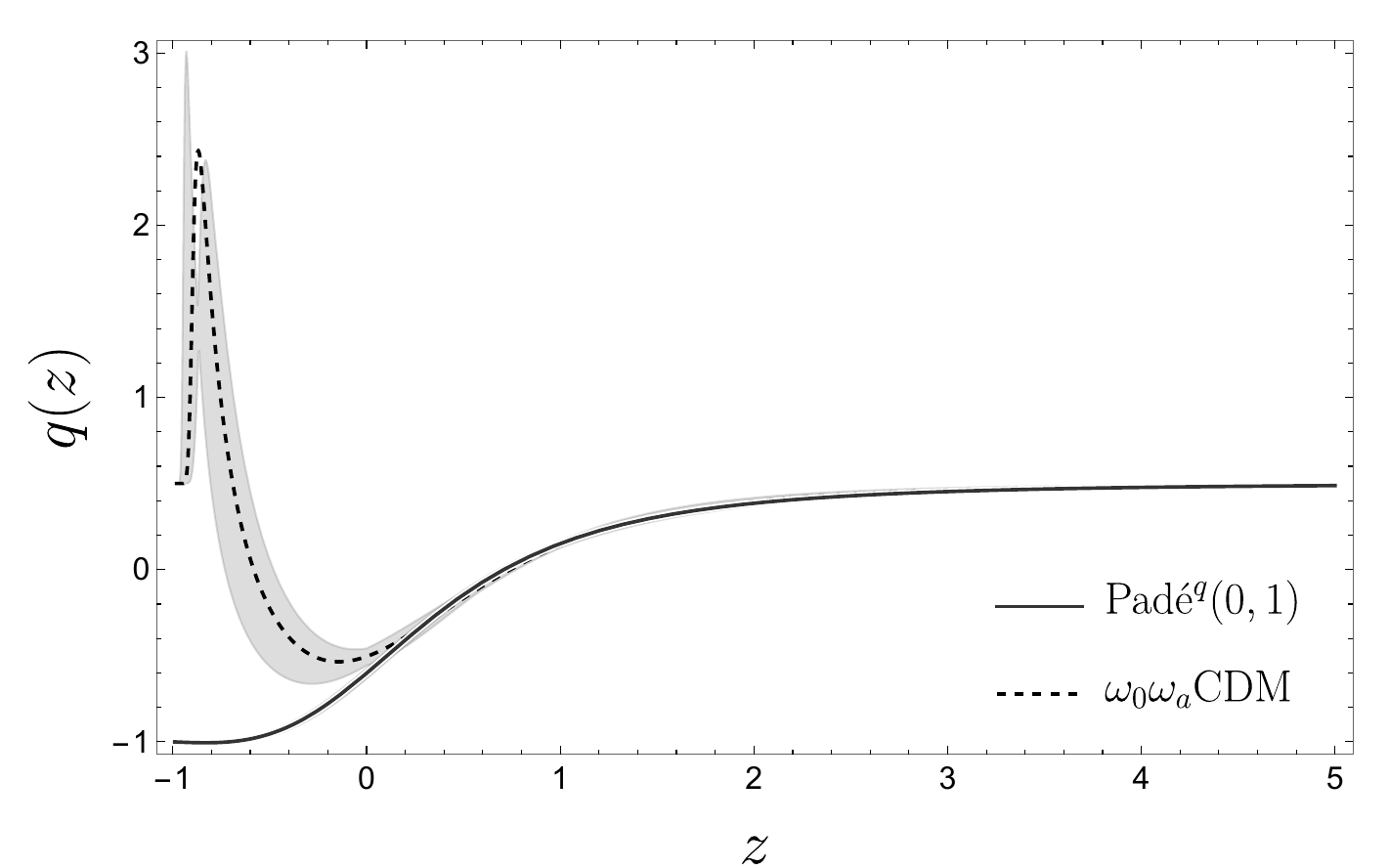}\hfill
     \includegraphics[width=0.5\textwidth]{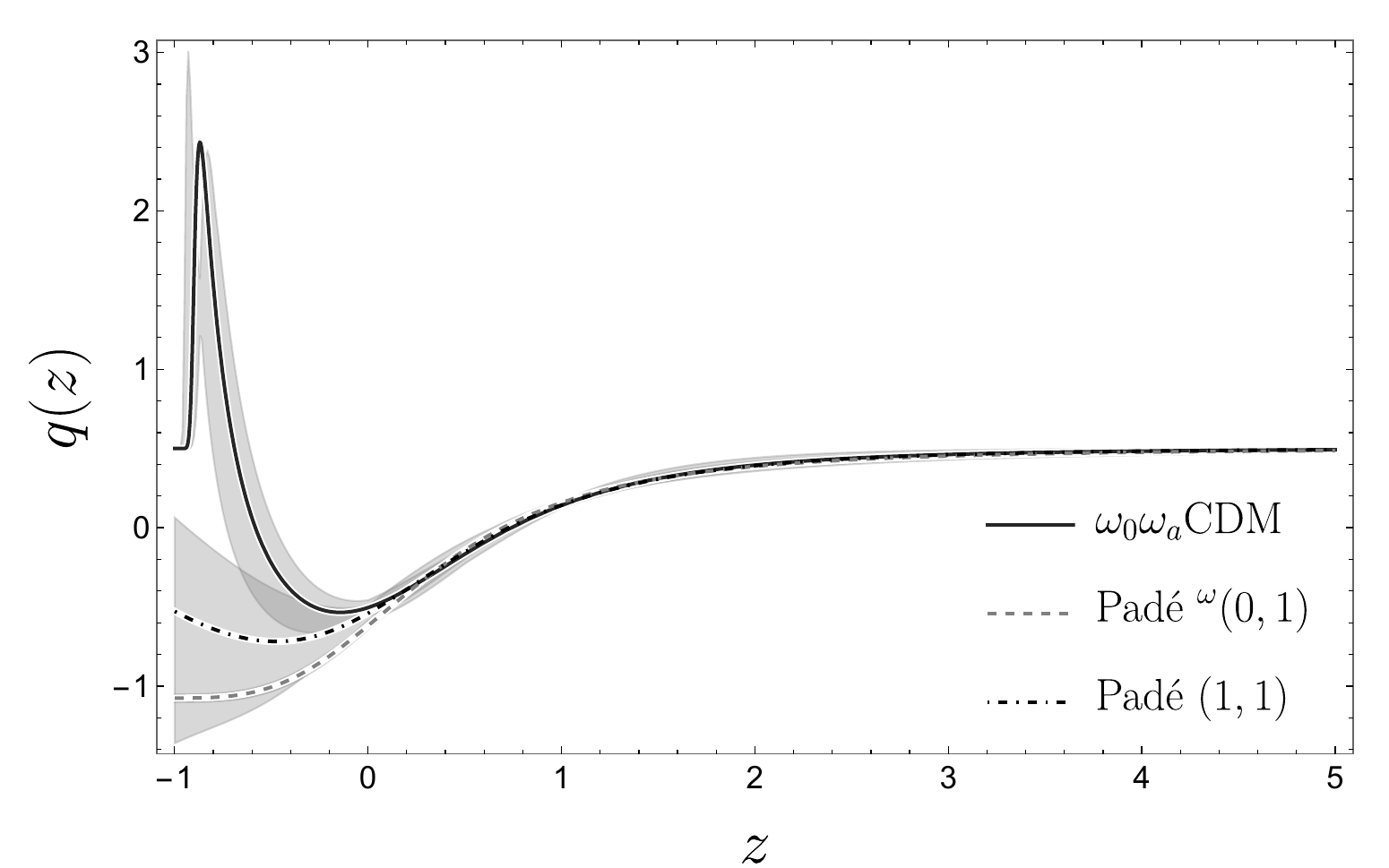}
    \caption{Reconstructed evolution of the deceleration parameter $q(z)$ based on the MCMC constraints for the proposed Padé dark energy parameterizations, compared to the $\Lambda$CDM and $\omega_0\omega_a$CDM models. The shaded bands indicate the $1\sigma$ uncertainty ranges around the mean.}
    \label{fig:decp}
\end{figure*}

Here, we present the evolution of the deceleration parameter $q(z)$, shown in Fig.~\ref{fig:decp}, evaluated using the best-fit parameters obtained from the MCMC analysis. The results indicate that all parameterizations behave consistently at high redshift, i.e., for $z \gg 1$, as expected by construction. 

However, the $\omega_0\omega_a$CDM model exhibits positive values of $q(z)$ as $z \rightarrow -1$, implying a future time behavior that differs from our initial expectations. Furthermore, model-dependent Padé parameterizations also show a slightly non-standard evolution in this regime as $z \rightarrow -1$, we obtain for Padé$^{\omega}$ (0,1) and Padé (1,1), $q\sim -1.07$ and $q\sim -0.5$, respectively.

\begin{thebibliography}{74}
\expandafter\ifx\csname natexlab\endcsname\relax\def\natexlab#1{#1}\fi
\expandafter\ifx\csname bibnamefont\endcsname\relax
  \def\bibnamefont#1{#1}\fi
\expandafter\ifx\csname bibfnamefont\endcsname\relax
  \def\bibfnamefont#1{#1}\fi
\expandafter\ifx\csname citenamefont\endcsname\relax
  \def\citenamefont#1{#1}\fi
\expandafter\ifx\csname url\endcsname\relax
  \def\url#1{\texttt{#1}}\fi
\expandafter\ifx\csname urlprefix\endcsname\relax\def\urlprefix{URL }\fi
\providecommand{\bibinfo}[2]{#2}
\providecommand{\eprint}[2][]{\url{#2}}

\bibitem[{\citenamefont{Padmanabhan}(2003)}]{Padmanabhan:2002ji}
\bibinfo{author}{\bibfnamefont{T.}~\bibnamefont{Padmanabhan}}, \bibinfo{journal}{Phys. Rept.} \textbf{\bibinfo{volume}{380}}, \bibinfo{pages}{235} (\bibinfo{year}{2003}), \eprint{hep-th/0212290}.

\bibitem[{\citenamefont{Peebles and Ratra}(2003)}]{Peebles:2002gy}
\bibinfo{author}{\bibfnamefont{P.~J.~E.} \bibnamefont{Peebles}} \bibnamefont{and} \bibinfo{author}{\bibfnamefont{B.}~\bibnamefont{Ratra}}, \bibinfo{journal}{Rev. Mod. Phys.} \textbf{\bibinfo{volume}{75}}, \bibinfo{pages}{559} (\bibinfo{year}{2003}), \eprint{astro-ph/0207347}.

\bibitem[{\citenamefont{Copeland et~al.}(2006)\citenamefont{Copeland, Sami, and Tsujikawa}}]{Copeland:2006wr}
\bibinfo{author}{\bibfnamefont{E.~J.} \bibnamefont{Copeland}}, \bibinfo{author}{\bibfnamefont{M.}~\bibnamefont{Sami}}, \bibnamefont{and} \bibinfo{author}{\bibfnamefont{S.}~\bibnamefont{Tsujikawa}}, \bibinfo{journal}{Int. J. Mod. Phys. D} \textbf{\bibinfo{volume}{15}}, \bibinfo{pages}{1753} (\bibinfo{year}{2006}), \eprint{hep-th/0603057}.

\bibitem[{\citenamefont{Perlmutter et~al.}(1998)}]{SupernovaCosmologyProject:1997zqe}
\bibinfo{author}{\bibfnamefont{S.}~\bibnamefont{Perlmutter}} \bibnamefont{et~al.} (\bibinfo{collaboration}{Supernova Cosmology Project}), \bibinfo{journal}{Nature} \textbf{\bibinfo{volume}{391}}, \bibinfo{pages}{51} (\bibinfo{year}{1998}), \eprint{astro-ph/9712212}.

\bibitem[{\citenamefont{Perlmutter et~al.}(1999)}]{SupernovaCosmologyProject:1998vns}
\bibinfo{author}{\bibfnamefont{S.}~\bibnamefont{Perlmutter}} \bibnamefont{et~al.} (\bibinfo{collaboration}{Supernova Cosmology Project}), \bibinfo{journal}{Astrophys. J.} \textbf{\bibinfo{volume}{517}}, \bibinfo{pages}{565} (\bibinfo{year}{1999}), \eprint{astro-ph/9812133}.

\bibitem[{\citenamefont{Riess et~al.}(1998)}]{SupernovaSearchTeam:1998fmf}
\bibinfo{author}{\bibfnamefont{A.~G.} \bibnamefont{Riess}} \bibnamefont{et~al.} (\bibinfo{collaboration}{Supernova Search Team}), \bibinfo{journal}{Astron. J.} \textbf{\bibinfo{volume}{116}}, \bibinfo{pages}{1009} (\bibinfo{year}{1998}), \eprint{astro-ph/9805201}.

\bibitem[{\citenamefont{Tonry et~al.}(2003)}]{SupernovaSearchTeam:2003cyd}
\bibinfo{author}{\bibfnamefont{J.~L.} \bibnamefont{Tonry}} \bibnamefont{et~al.} (\bibinfo{collaboration}{Supernova Search Team}), \bibinfo{journal}{Astrophys. J.} \textbf{\bibinfo{volume}{594}}, \bibinfo{pages}{1} (\bibinfo{year}{2003}), \eprint{astro-ph/0305008}.

\bibitem[{\citenamefont{Huterer and Shafer}(2018)}]{Huterer:2017buf}
\bibinfo{author}{\bibfnamefont{D.}~\bibnamefont{Huterer}} \bibnamefont{and} \bibinfo{author}{\bibfnamefont{D.~L.} \bibnamefont{Shafer}}, \bibinfo{journal}{Rept. Prog. Phys.} \textbf{\bibinfo{volume}{81}}, \bibinfo{pages}{016901} (\bibinfo{year}{2018}), \eprint{1709.01091}.

\bibitem[{\citenamefont{Aghanim et~al.}(2020)}]{Planck:2018vyg}
\bibinfo{author}{\bibfnamefont{N.}~\bibnamefont{Aghanim}} \bibnamefont{et~al.} (\bibinfo{collaboration}{Planck}), \bibinfo{journal}{Astron. Astrophys.} \textbf{\bibinfo{volume}{641}}, \bibinfo{pages}{A6} (\bibinfo{year}{2020}), \bibinfo{note}{[Erratum: Astron.Astrophys. 652, C4 (2021)]}, \eprint{1807.06209}.

\bibitem[{\citenamefont{Weinberg}(1989)}]{Weinberg:1988cp}
\bibinfo{author}{\bibfnamefont{S.}~\bibnamefont{Weinberg}}, \bibinfo{journal}{Rev. Mod. Phys.} \textbf{\bibinfo{volume}{61}}, \bibinfo{pages}{1} (\bibinfo{year}{1989}).

\bibitem[{\citenamefont{Martin}(2012)}]{Martin:2012bt}
\bibinfo{author}{\bibfnamefont{J.}~\bibnamefont{Martin}}, \bibinfo{journal}{Comptes Rendus Physique} \textbf{\bibinfo{volume}{13}}, \bibinfo{pages}{566} (\bibinfo{year}{2012}), \eprint{1205.3365}.

\bibitem[{\citenamefont{Sola~Peracaula}(2022)}]{SolaPeracaula:2022hpd}
\bibinfo{author}{\bibfnamefont{J.}~\bibnamefont{Sola~Peracaula}}, \bibinfo{journal}{Phil. Trans. Roy. Soc. Lond. A} \textbf{\bibinfo{volume}{380}}, \bibinfo{pages}{20210182} (\bibinfo{year}{2022}), \eprint{2203.13757}.

\bibitem[{\citenamefont{Belfiglio et~al.}(2024)\citenamefont{Belfiglio, Carloni, and Luongo}}]{Belfiglio:2023rxb}
\bibinfo{author}{\bibfnamefont{A.}~\bibnamefont{Belfiglio}}, \bibinfo{author}{\bibfnamefont{Y.}~\bibnamefont{Carloni}}, \bibnamefont{and} \bibinfo{author}{\bibfnamefont{O.}~\bibnamefont{Luongo}}, \bibinfo{journal}{Phys. Dark Univ.} \textbf{\bibinfo{volume}{44}}, \bibinfo{pages}{101458} (\bibinfo{year}{2024}), \eprint{2307.04739}.

\bibitem[{\citenamefont{Kamionkowski and Riess}(2023)}]{Kamionkowski:2022pkx}
\bibinfo{author}{\bibfnamefont{M.}~\bibnamefont{Kamionkowski}} \bibnamefont{and} \bibinfo{author}{\bibfnamefont{A.~G.} \bibnamefont{Riess}}, \bibinfo{journal}{Ann. Rev. Nucl. Part. Sci.} \textbf{\bibinfo{volume}{73}}, \bibinfo{pages}{153} (\bibinfo{year}{2023}), \eprint{2211.04492}.

\bibitem[{\citenamefont{Sch{\"o}neberg et~al.}(2022)\citenamefont{Sch{\"o}neberg, Franco~Abell{\'a}n, P{\'e}rez~S{\'a}nchez, Witte, Poulin, and Lesgourgues}}]{Schoneberg:2021qvd}
\bibinfo{author}{\bibfnamefont{N.}~\bibnamefont{Sch{\"o}neberg}}, \bibinfo{author}{\bibfnamefont{G.}~\bibnamefont{Franco~Abell{\'a}n}}, \bibinfo{author}{\bibfnamefont{A.}~\bibnamefont{P{\'e}rez~S{\'a}nchez}}, \bibinfo{author}{\bibfnamefont{S.~J.} \bibnamefont{Witte}}, \bibinfo{author}{\bibfnamefont{V.}~\bibnamefont{Poulin}}, \bibnamefont{and} \bibinfo{author}{\bibfnamefont{J.}~\bibnamefont{Lesgourgues}}, \bibinfo{journal}{Phys. Rept.} \textbf{\bibinfo{volume}{984}}, \bibinfo{pages}{1} (\bibinfo{year}{2022}), \eprint{2107.10291}.

\bibitem[{\citenamefont{Poulin et~al.}(2023)\citenamefont{Poulin, Smith, and Karwal}}]{Poulin:2023lkg}
\bibinfo{author}{\bibfnamefont{V.}~\bibnamefont{Poulin}}, \bibinfo{author}{\bibfnamefont{T.~L.} \bibnamefont{Smith}}, \bibnamefont{and} \bibinfo{author}{\bibfnamefont{T.}~\bibnamefont{Karwal}}, \bibinfo{journal}{Phys. Dark Univ.} \textbf{\bibinfo{volume}{42}}, \bibinfo{pages}{101348} (\bibinfo{year}{2023}), \eprint{2302.09032}.

\bibitem[{\citenamefont{Abdalla et~al.}(2022)}]{Abdalla:2022yfr}
\bibinfo{author}{\bibfnamefont{E.}~\bibnamefont{Abdalla}} \bibnamefont{et~al.}, \bibinfo{journal}{JHEAp} \textbf{\bibinfo{volume}{34}}, \bibinfo{pages}{49} (\bibinfo{year}{2022}), \eprint{2203.06142}.

\bibitem[{\citenamefont{Di~Valentino et~al.}(2021)}]{DiValentino:2020vvd}
\bibinfo{author}{\bibfnamefont{E.}~\bibnamefont{Di~Valentino}} \bibnamefont{et~al.}, \bibinfo{journal}{Astropart. Phys.} \textbf{\bibinfo{volume}{131}}, \bibinfo{pages}{102604} (\bibinfo{year}{2021}), \eprint{2008.11285}.

\bibitem[{\citenamefont{Carloni et~al.}(2025{\natexlab{a}})\citenamefont{Carloni, Luongo, and Muccino}}]{Carloni:2025jlk}
\bibinfo{author}{\bibfnamefont{Y.}~\bibnamefont{Carloni}}, \bibinfo{author}{\bibfnamefont{O.}~\bibnamefont{Luongo}}, \bibnamefont{and} \bibinfo{author}{\bibfnamefont{M.}~\bibnamefont{Muccino}} (\bibinfo{year}{2025}{\natexlab{a}}), \eprint{2506.11531}.

\bibitem[{\citenamefont{Adame et~al.}(2025)}]{DESI:2024mwx}
\bibinfo{author}{\bibfnamefont{A.~G.} \bibnamefont{Adame}} \bibnamefont{et~al.} (\bibinfo{collaboration}{DESI}), \bibinfo{journal}{JCAP} \textbf{\bibinfo{volume}{02}}, \bibinfo{pages}{021} (\bibinfo{year}{2025}), \eprint{2404.03002}.

\bibitem[{\citenamefont{Cort{\^e}s and Liddle}(2024)}]{Cortes:2024lgw}
\bibinfo{author}{\bibfnamefont{M.}~\bibnamefont{Cort{\^e}s}} \bibnamefont{and} \bibinfo{author}{\bibfnamefont{A.~R.} \bibnamefont{Liddle}}, \bibinfo{journal}{JCAP} \textbf{\bibinfo{volume}{12}}, \bibinfo{pages}{007} (\bibinfo{year}{2024}), \eprint{2404.08056}.

\bibitem[{\citenamefont{Giar{\`e} et~al.}(2024)\citenamefont{Giar{\`e}, Najafi, Pan, Di~Valentino, and Firouzjaee}}]{Giare:2024gpk}
\bibinfo{author}{\bibfnamefont{W.}~\bibnamefont{Giar{\`e}}}, \bibinfo{author}{\bibfnamefont{M.}~\bibnamefont{Najafi}}, \bibinfo{author}{\bibfnamefont{S.}~\bibnamefont{Pan}}, \bibinfo{author}{\bibfnamefont{E.}~\bibnamefont{Di~Valentino}}, \bibnamefont{and} \bibinfo{author}{\bibfnamefont{J.~T.} \bibnamefont{Firouzjaee}}, \bibinfo{journal}{JCAP} \textbf{\bibinfo{volume}{10}}, \bibinfo{pages}{035} (\bibinfo{year}{2024}), \eprint{2407.16689}.

\bibitem[{\citenamefont{Abdul~Karim et~al.}(2025)}]{DESI:2025zgx}
\bibinfo{author}{\bibfnamefont{M.}~\bibnamefont{Abdul~Karim}} \bibnamefont{et~al.} (\bibinfo{collaboration}{DESI}) (\bibinfo{year}{2025}), \eprint{2503.14738}.

\bibitem[{\citenamefont{Colg{\'a}in et~al.}(2026)\citenamefont{Colg{\'a}in, Dainotti, Capozziello, Pourojaghi, Sheikh-Jabbari, and Stojkovic}}]{Colgain:2024xqj}
\bibinfo{author}{\bibfnamefont{E.~{\'O}.} \bibnamefont{Colg{\'a}in}}, \bibinfo{author}{\bibfnamefont{M.~G.} \bibnamefont{Dainotti}}, \bibinfo{author}{\bibfnamefont{S.}~\bibnamefont{Capozziello}}, \bibinfo{author}{\bibfnamefont{S.}~\bibnamefont{Pourojaghi}}, \bibinfo{author}{\bibfnamefont{M.~M.} \bibnamefont{Sheikh-Jabbari}}, \bibnamefont{and} \bibinfo{author}{\bibfnamefont{D.}~\bibnamefont{Stojkovic}}, \bibinfo{journal}{JHEAp} \textbf{\bibinfo{volume}{49}}, \bibinfo{pages}{100428} (\bibinfo{year}{2026}), \eprint{2404.08633}.

\bibitem[{\citenamefont{Luongo and Muccino}(2024)}]{Luongo:2024fww}
\bibinfo{author}{\bibfnamefont{O.}~\bibnamefont{Luongo}} \bibnamefont{and} \bibinfo{author}{\bibfnamefont{M.}~\bibnamefont{Muccino}}, \bibinfo{journal}{Astron. Astrophys.} \textbf{\bibinfo{volume}{690}}, \bibinfo{pages}{A40} (\bibinfo{year}{2024}), \eprint{2404.07070}.

\bibitem[{\citenamefont{Carloni et~al.}(2025{\natexlab{b}})\citenamefont{Carloni, Luongo, and Muccino}}]{Carloni:2024zpl}
\bibinfo{author}{\bibfnamefont{Y.}~\bibnamefont{Carloni}}, \bibinfo{author}{\bibfnamefont{O.}~\bibnamefont{Luongo}}, \bibnamefont{and} \bibinfo{author}{\bibfnamefont{M.}~\bibnamefont{Muccino}}, \bibinfo{journal}{Phys. Rev. D} \textbf{\bibinfo{volume}{111}}, \bibinfo{pages}{023512} (\bibinfo{year}{2025}{\natexlab{b}}), \eprint{2404.12068}.

\bibitem[{\citenamefont{Colg{\'a}in et~al.}(2025{\natexlab{a}})\citenamefont{Colg{\'a}in, Pourojaghi, Sheikh-Jabbari, and Yin}}]{Colgain:2025nzf}
\bibinfo{author}{\bibfnamefont{E.~{\'O}.} \bibnamefont{Colg{\'a}in}}, \bibinfo{author}{\bibfnamefont{S.}~\bibnamefont{Pourojaghi}}, \bibinfo{author}{\bibfnamefont{M.~M.} \bibnamefont{Sheikh-Jabbari}}, \bibnamefont{and} \bibinfo{author}{\bibfnamefont{L.}~\bibnamefont{Yin}} (\bibinfo{year}{2025}{\natexlab{a}}), \eprint{2504.04417}.

\bibitem[{\citenamefont{Liu et~al.}(2025)\citenamefont{Liu, Li, Xu, Biesiada, and Wang}}]{Liu:2025myr}
\bibinfo{author}{\bibfnamefont{T.}~\bibnamefont{Liu}}, \bibinfo{author}{\bibfnamefont{X.}~\bibnamefont{Li}}, \bibinfo{author}{\bibfnamefont{T.}~\bibnamefont{Xu}}, \bibinfo{author}{\bibfnamefont{M.}~\bibnamefont{Biesiada}}, \bibnamefont{and} \bibinfo{author}{\bibfnamefont{J.}~\bibnamefont{Wang}} (\bibinfo{year}{2025}), \eprint{2507.04265}.

\bibitem[{\citenamefont{Gruber and Luongo}(2014)}]{Gruber:2013wua}
\bibinfo{author}{\bibfnamefont{C.}~\bibnamefont{Gruber}} \bibnamefont{and} \bibinfo{author}{\bibfnamefont{O.}~\bibnamefont{Luongo}}, \bibinfo{journal}{Phys. Rev. D} \textbf{\bibinfo{volume}{89}}, \bibinfo{pages}{103506} (\bibinfo{year}{2014}), \eprint{1309.3215}.

\bibitem[{\citenamefont{Aviles et~al.}(2014)\citenamefont{Aviles, Bravetti, Capozziello, and Luongo}}]{Aviles:2014rma}
\bibinfo{author}{\bibfnamefont{A.}~\bibnamefont{Aviles}}, \bibinfo{author}{\bibfnamefont{A.}~\bibnamefont{Bravetti}}, \bibinfo{author}{\bibfnamefont{S.}~\bibnamefont{Capozziello}}, \bibnamefont{and} \bibinfo{author}{\bibfnamefont{O.}~\bibnamefont{Luongo}}, \bibinfo{journal}{Phys. Rev. D} \textbf{\bibinfo{volume}{90}}, \bibinfo{pages}{043531} (\bibinfo{year}{2014}), \eprint{1405.6935}.

\bibitem[{\citenamefont{Rezaei et~al.}(2017)\citenamefont{Rezaei, Malekjani, Basilakos, Mehrabi, and Mota}}]{Rezaei:2017yyj}
\bibinfo{author}{\bibfnamefont{M.}~\bibnamefont{Rezaei}}, \bibinfo{author}{\bibfnamefont{M.}~\bibnamefont{Malekjani}}, \bibinfo{author}{\bibfnamefont{S.}~\bibnamefont{Basilakos}}, \bibinfo{author}{\bibfnamefont{A.}~\bibnamefont{Mehrabi}}, \bibnamefont{and} \bibinfo{author}{\bibfnamefont{D.~F.} \bibnamefont{Mota}}, \bibinfo{journal}{Astrophys. J.} \textbf{\bibinfo{volume}{843}}, \bibinfo{pages}{65} (\bibinfo{year}{2017}), \eprint{1706.02537}.

\bibitem[{\citenamefont{Capozziello et~al.}(2018)\citenamefont{Capozziello, D'Agostino, and Luongo}}]{Capozziello:2017ddd}
\bibinfo{author}{\bibfnamefont{S.}~\bibnamefont{Capozziello}}, \bibinfo{author}{\bibfnamefont{R.}~\bibnamefont{D'Agostino}}, \bibnamefont{and} \bibinfo{author}{\bibfnamefont{O.}~\bibnamefont{Luongo}}, \bibinfo{journal}{JCAP} \textbf{\bibinfo{volume}{05}}, \bibinfo{pages}{008} (\bibinfo{year}{2018}), \eprint{1709.08407}.

\bibitem[{\citenamefont{Capozziello et~al.}(2020)\citenamefont{Capozziello, D'Agostino, and Luongo}}]{Capozziello:2020ctn}
\bibinfo{author}{\bibfnamefont{S.}~\bibnamefont{Capozziello}}, \bibinfo{author}{\bibfnamefont{R.}~\bibnamefont{D'Agostino}}, \bibnamefont{and} \bibinfo{author}{\bibfnamefont{O.}~\bibnamefont{Luongo}}, \bibinfo{journal}{Mon. Not. Roy. Astron. Soc.} \textbf{\bibinfo{volume}{494}}, \bibinfo{pages}{2576} (\bibinfo{year}{2020}), \eprint{2003.09341}.

\bibitem[{\citenamefont{Capozziello et~al.}(2022)\citenamefont{Capozziello, D'Agostino, and Luongo}}]{Capozziello:2022jbw}
\bibinfo{author}{\bibfnamefont{S.}~\bibnamefont{Capozziello}}, \bibinfo{author}{\bibfnamefont{R.}~\bibnamefont{D'Agostino}}, \bibnamefont{and} \bibinfo{author}{\bibfnamefont{O.}~\bibnamefont{Luongo}}, \bibinfo{journal}{Phys. Dark Univ.} \textbf{\bibinfo{volume}{36}}, \bibinfo{pages}{101045} (\bibinfo{year}{2022}), \eprint{2202.03300}.

\bibitem[{\citenamefont{Armendariz-Picon et~al.}(2000)\citenamefont{Armendariz-Picon, Mukhanov, and Steinhardt}}]{Armendariz-Picon:2000nqq}
\bibinfo{author}{\bibfnamefont{C.}~\bibnamefont{Armendariz-Picon}}, \bibinfo{author}{\bibfnamefont{V.~F.} \bibnamefont{Mukhanov}}, \bibnamefont{and} \bibinfo{author}{\bibfnamefont{P.~J.} \bibnamefont{Steinhardt}}, \bibinfo{journal}{Phys. Rev. Lett.} \textbf{\bibinfo{volume}{85}}, \bibinfo{pages}{4438} (\bibinfo{year}{2000}), \eprint{astro-ph/0004134}.

\bibitem[{\citenamefont{Kunz}(2012)}]{Kunz:2012aw}
\bibinfo{author}{\bibfnamefont{M.}~\bibnamefont{Kunz}}, \bibinfo{journal}{Comptes Rendus Physique} \textbf{\bibinfo{volume}{13}}, \bibinfo{pages}{539} (\bibinfo{year}{2012}), \eprint{1204.5482}.

\bibitem[{\citenamefont{Tsujikawa}(2013)}]{Tsujikawa:2013fta}
\bibinfo{author}{\bibfnamefont{S.}~\bibnamefont{Tsujikawa}}, \bibinfo{journal}{Class. Quant. Grav.} \textbf{\bibinfo{volume}{30}}, \bibinfo{pages}{214003} (\bibinfo{year}{2013}), \eprint{1304.1961}.

\bibitem[{\citenamefont{Wolf et~al.}(2024)\citenamefont{Wolf, Garc{\'\i}a-Garc{\'\i}a, Bartlett, and Ferreira}}]{Wolf:2024eph}
\bibinfo{author}{\bibfnamefont{W.~J.} \bibnamefont{Wolf}}, \bibinfo{author}{\bibfnamefont{C.}~\bibnamefont{Garc{\'\i}a-Garc{\'\i}a}}, \bibinfo{author}{\bibfnamefont{D.~J.} \bibnamefont{Bartlett}}, \bibnamefont{and} \bibinfo{author}{\bibfnamefont{P.~G.} \bibnamefont{Ferreira}}, \bibinfo{journal}{Phys. Rev. D} \textbf{\bibinfo{volume}{110}}, \bibinfo{pages}{083528} (\bibinfo{year}{2024}), \eprint{2408.17318}.

\bibitem[{\citenamefont{Wolf et~al.}(2025)\citenamefont{Wolf, Garc{\'\i}a-Garc{\'\i}a, and Ferreira}}]{Wolf:2025jlc}
\bibinfo{author}{\bibfnamefont{W.~J.} \bibnamefont{Wolf}}, \bibinfo{author}{\bibfnamefont{C.}~\bibnamefont{Garc{\'\i}a-Garc{\'\i}a}}, \bibnamefont{and} \bibinfo{author}{\bibfnamefont{P.~G.} \bibnamefont{Ferreira}}, \bibinfo{journal}{JCAP} \textbf{\bibinfo{volume}{05}}, \bibinfo{pages}{034} (\bibinfo{year}{2025}), \eprint{2502.04929}.

\bibitem[{\citenamefont{{Di Valentino} et~al.}(2025)\citenamefont{{Di Valentino}, {Said}, {Riess}, {Pollo}, {Poulin}, {G{\'o}mez-Valent}, {Weltman}, {Palmese}, {Huang}, {van de Bruck} et~al.}}]{2025PDU....4901965D}
\bibinfo{author}{\bibfnamefont{E.}~\bibnamefont{{Di Valentino}}}, \bibinfo{author}{\bibfnamefont{J.~L.} \bibnamefont{{Said}}}, \bibinfo{author}{\bibfnamefont{A.}~\bibnamefont{{Riess}}}, \bibinfo{author}{\bibfnamefont{A.}~\bibnamefont{{Pollo}}}, \bibinfo{author}{\bibfnamefont{V.}~\bibnamefont{{Poulin}}}, \bibinfo{author}{\bibfnamefont{A.}~\bibnamefont{{G{\'o}mez-Valent}}}, \bibinfo{author}{\bibfnamefont{A.}~\bibnamefont{{Weltman}}}, \bibinfo{author}{\bibfnamefont{A.}~\bibnamefont{{Palmese}}}, \bibinfo{author}{\bibfnamefont{C.~D.} \bibnamefont{{Huang}}}, \bibinfo{author}{\bibfnamefont{C.}~\bibnamefont{{van de Bruck}}}, \bibnamefont{et~al.}, \bibinfo{journal}{Physics of the Dark Universe} \textbf{\bibinfo{volume}{49}}, \bibinfo{eid}{101965} (\bibinfo{year}{2025}), \eprint{2504.01669}.

\bibitem[{\citenamefont{Starobinsky}(1998)}]{Starobinsky:1998fr}
\bibinfo{author}{\bibfnamefont{A.~A.} \bibnamefont{Starobinsky}}, \bibinfo{journal}{JETP Lett.} \textbf{\bibinfo{volume}{68}}, \bibinfo{pages}{757} (\bibinfo{year}{1998}), \eprint{astro-ph/9810431}.

\bibitem[{\citenamefont{Huterer and Turner}(1999)}]{Huterer:1998qv}
\bibinfo{author}{\bibfnamefont{D.}~\bibnamefont{Huterer}} \bibnamefont{and} \bibinfo{author}{\bibfnamefont{M.~S.} \bibnamefont{Turner}}, \bibinfo{journal}{Phys. Rev. D} \textbf{\bibinfo{volume}{60}}, \bibinfo{pages}{081301} (\bibinfo{year}{1999}), \eprint{astro-ph/9808133}.

\bibitem[{\citenamefont{Sahni and Starobinsky}(2006)}]{Sahni:2006pa}
\bibinfo{author}{\bibfnamefont{V.}~\bibnamefont{Sahni}} \bibnamefont{and} \bibinfo{author}{\bibfnamefont{A.}~\bibnamefont{Starobinsky}}, \bibinfo{journal}{Int. J. Mod. Phys. D} \textbf{\bibinfo{volume}{15}}, \bibinfo{pages}{2105} (\bibinfo{year}{2006}), \eprint{astro-ph/0610026}.

\bibitem[{\citenamefont{Cattoen and Visser}(2007)}]{Cattoen:2007sk}
\bibinfo{author}{\bibfnamefont{C.}~\bibnamefont{Cattoen}} \bibnamefont{and} \bibinfo{author}{\bibfnamefont{M.}~\bibnamefont{Visser}}, \bibinfo{journal}{Class. Quant. Grav.} \textbf{\bibinfo{volume}{24}}, \bibinfo{pages}{5985} (\bibinfo{year}{2007}), \eprint{0710.1887}.

\bibitem[{\citenamefont{Bamba et~al.}(2012)\citenamefont{Bamba, Capozziello, Nojiri, and Odintsov}}]{Bamba:2012cp}
\bibinfo{author}{\bibfnamefont{K.}~\bibnamefont{Bamba}}, \bibinfo{author}{\bibfnamefont{S.}~\bibnamefont{Capozziello}}, \bibinfo{author}{\bibfnamefont{S.}~\bibnamefont{Nojiri}}, \bibnamefont{and} \bibinfo{author}{\bibfnamefont{S.~D.} \bibnamefont{Odintsov}}, \bibinfo{journal}{Astrophys. Space Sci.} \textbf{\bibinfo{volume}{342}}, \bibinfo{pages}{155} (\bibinfo{year}{2012}), \eprint{1205.3421}.

\bibitem[{\citenamefont{{Ratra} and {Peebles}}(1988)}]{RatraPeebles1988}
\bibinfo{author}{\bibfnamefont{B.}~\bibnamefont{{Ratra}}} \bibnamefont{and} \bibinfo{author}{\bibfnamefont{P.~J.~E.} \bibnamefont{{Peebles}}}, \bibinfo{journal}{\prd} \textbf{\bibinfo{volume}{37}}, \bibinfo{pages}{3406} (\bibinfo{year}{1988}).

\bibitem[{\citenamefont{Efstathiou and Gratton}(2020)}]{Efstathiou:2020wem}
\bibinfo{author}{\bibfnamefont{G.}~\bibnamefont{Efstathiou}} \bibnamefont{and} \bibinfo{author}{\bibfnamefont{S.}~\bibnamefont{Gratton}}, \bibinfo{journal}{Mon. Not. Roy. Astron. Soc.} \textbf{\bibinfo{volume}{496}}, \bibinfo{pages}{L91} (\bibinfo{year}{2020}), \eprint{2002.06892}.

\bibitem[{\citenamefont{Aviles et~al.}(2012)\citenamefont{Aviles, Gruber, Luongo, and Quevedo}}]{Aviles:2012ay}
\bibinfo{author}{\bibfnamefont{A.}~\bibnamefont{Aviles}}, \bibinfo{author}{\bibfnamefont{C.}~\bibnamefont{Gruber}}, \bibinfo{author}{\bibfnamefont{O.}~\bibnamefont{Luongo}}, \bibnamefont{and} \bibinfo{author}{\bibfnamefont{H.}~\bibnamefont{Quevedo}}, \bibinfo{journal}{Phys. Rev. D} \textbf{\bibinfo{volume}{86}}, \bibinfo{pages}{123516} (\bibinfo{year}{2012}), \eprint{1204.2007}.

\bibitem[{\citenamefont{Hu and Wang}(2022)}]{Hu:2022udt}
\bibinfo{author}{\bibfnamefont{J.~P.} \bibnamefont{Hu}} \bibnamefont{and} \bibinfo{author}{\bibfnamefont{F.~Y.} \bibnamefont{Wang}}, \bibinfo{journal}{Astron. Astrophys.} \textbf{\bibinfo{volume}{661}}, \bibinfo{pages}{A71} (\bibinfo{year}{2022}), \eprint{2202.09075}.

\bibitem[{\citenamefont{Chevallier and Polarski}(2001)}]{Chevallier:2000qy}
\bibinfo{author}{\bibfnamefont{M.}~\bibnamefont{Chevallier}} \bibnamefont{and} \bibinfo{author}{\bibfnamefont{D.}~\bibnamefont{Polarski}}, \bibinfo{journal}{Int. J. Mod. Phys. D} \textbf{\bibinfo{volume}{10}}, \bibinfo{pages}{213} (\bibinfo{year}{2001}), \eprint{gr-qc/0009008}.

\bibitem[{\citenamefont{Linder}(2003)}]{Linder:2002et}
\bibinfo{author}{\bibfnamefont{E.~V.} \bibnamefont{Linder}}, \bibinfo{journal}{Phys. Rev. Lett.} \textbf{\bibinfo{volume}{90}}, \bibinfo{pages}{091301} (\bibinfo{year}{2003}), \eprint{astro-ph/0208512}.

\bibitem[{\citenamefont{Scolnic et~al.}(2022)}]{Scolnic:2021amr}
\bibinfo{author}{\bibfnamefont{D.}~\bibnamefont{Scolnic}} \bibnamefont{et~al.}, \bibinfo{journal}{Astrophys. J.} \textbf{\bibinfo{volume}{938}}, \bibinfo{pages}{113} (\bibinfo{year}{2022}), \eprint{2112.03863}.

\bibitem[{\citenamefont{Brout et~al.}(2022)}]{Brout:2022vxf}
\bibinfo{author}{\bibfnamefont{D.}~\bibnamefont{Brout}} \bibnamefont{et~al.}, \bibinfo{journal}{Astrophys. J.} \textbf{\bibinfo{volume}{938}}, \bibinfo{pages}{110} (\bibinfo{year}{2022}), \eprint{2202.04077}.

\bibitem[{\citenamefont{Conley et~al.}(2011)}]{SNLS:2011lii}
\bibinfo{author}{\bibfnamefont{A.}~\bibnamefont{Conley}} \bibnamefont{et~al.} (\bibinfo{collaboration}{SNLS}), \bibinfo{journal}{Astrophys. J. Suppl.} \textbf{\bibinfo{volume}{192}}, \bibinfo{pages}{1} (\bibinfo{year}{2011}), \eprint{1104.1443}.

\bibitem[{\citenamefont{Chen et~al.}(2019)\citenamefont{Chen, Huang, and Wang}}]{Chen:2018dbv}
\bibinfo{author}{\bibfnamefont{L.}~\bibnamefont{Chen}}, \bibinfo{author}{\bibfnamefont{Q.-G.} \bibnamefont{Huang}}, \bibnamefont{and} \bibinfo{author}{\bibfnamefont{K.}~\bibnamefont{Wang}}, \bibinfo{journal}{JCAP} \textbf{\bibinfo{volume}{02}}, \bibinfo{pages}{028} (\bibinfo{year}{2019}), \eprint{1808.05724}.

\bibitem[{\citenamefont{Jimenez and Loeb}(2002)}]{Jimenez:2001gg}
\bibinfo{author}{\bibfnamefont{R.}~\bibnamefont{Jimenez}} \bibnamefont{and} \bibinfo{author}{\bibfnamefont{A.}~\bibnamefont{Loeb}}, \bibinfo{journal}{Astrophys. J.} \textbf{\bibinfo{volume}{573}}, \bibinfo{pages}{37} (\bibinfo{year}{2002}), \eprint{astro-ph/0106145}.

\bibitem[{\citenamefont{Zhang et~al.}(2014)\citenamefont{Zhang, Zhang, Yuan, Zhang, and Sun}}]{Zhang:2012mp}
\bibinfo{author}{\bibfnamefont{C.}~\bibnamefont{Zhang}}, \bibinfo{author}{\bibfnamefont{H.}~\bibnamefont{Zhang}}, \bibinfo{author}{\bibfnamefont{S.}~\bibnamefont{Yuan}}, \bibinfo{author}{\bibfnamefont{T.-J.} \bibnamefont{Zhang}}, \bibnamefont{and} \bibinfo{author}{\bibfnamefont{Y.-C.} \bibnamefont{Sun}}, \bibinfo{journal}{Res. Astron. Astrophys.} \textbf{\bibinfo{volume}{14}}, \bibinfo{pages}{1221} (\bibinfo{year}{2014}), \eprint{1207.4541}.

\bibitem[{\citenamefont{Simon et~al.}(2005)\citenamefont{Simon, Verde, and Jimenez}}]{Simon:2004tf}
\bibinfo{author}{\bibfnamefont{J.}~\bibnamefont{Simon}}, \bibinfo{author}{\bibfnamefont{L.}~\bibnamefont{Verde}}, \bibnamefont{and} \bibinfo{author}{\bibfnamefont{R.}~\bibnamefont{Jimenez}}, \bibinfo{journal}{Phys. Rev. D} \textbf{\bibinfo{volume}{71}}, \bibinfo{pages}{123001} (\bibinfo{year}{2005}), \eprint{astro-ph/0412269}.

\bibitem[{\citenamefont{Moresco et~al.}(2012)}]{Moresco:2012jh}
\bibinfo{author}{\bibfnamefont{M.}~\bibnamefont{Moresco}} \bibnamefont{et~al.}, \bibinfo{journal}{JCAP} \textbf{\bibinfo{volume}{08}}, \bibinfo{pages}{006} (\bibinfo{year}{2012}), \eprint{1201.3609}.

\bibitem[{\citenamefont{Moresco et~al.}(2016)\citenamefont{Moresco, Pozzetti, Cimatti, Jimenez, Maraston, Verde, Thomas, Citro, Tojeiro, and Wilkinson}}]{Moresco:2016mzx}
\bibinfo{author}{\bibfnamefont{M.}~\bibnamefont{Moresco}}, \bibinfo{author}{\bibfnamefont{L.}~\bibnamefont{Pozzetti}}, \bibinfo{author}{\bibfnamefont{A.}~\bibnamefont{Cimatti}}, \bibinfo{author}{\bibfnamefont{R.}~\bibnamefont{Jimenez}}, \bibinfo{author}{\bibfnamefont{C.}~\bibnamefont{Maraston}}, \bibinfo{author}{\bibfnamefont{L.}~\bibnamefont{Verde}}, \bibinfo{author}{\bibfnamefont{D.}~\bibnamefont{Thomas}}, \bibinfo{author}{\bibfnamefont{A.}~\bibnamefont{Citro}}, \bibinfo{author}{\bibfnamefont{R.}~\bibnamefont{Tojeiro}}, \bibnamefont{and} \bibinfo{author}{\bibfnamefont{D.}~\bibnamefont{Wilkinson}}, \bibinfo{journal}{JCAP} \textbf{\bibinfo{volume}{05}}, \bibinfo{pages}{014} (\bibinfo{year}{2016}), \eprint{1601.01701}.

\bibitem[{\citenamefont{Ratsimbazafy et~al.}(2017)\citenamefont{Ratsimbazafy, Loubser, Crawford, Cress, Bassett, Nichol, and V{\"a}is{\"a}nen}}]{Ratsimbazafy:2017vga}
\bibinfo{author}{\bibfnamefont{A.~L.} \bibnamefont{Ratsimbazafy}}, \bibinfo{author}{\bibfnamefont{S.~I.} \bibnamefont{Loubser}}, \bibinfo{author}{\bibfnamefont{S.~M.} \bibnamefont{Crawford}}, \bibinfo{author}{\bibfnamefont{C.~M.} \bibnamefont{Cress}}, \bibinfo{author}{\bibfnamefont{B.~A.} \bibnamefont{Bassett}}, \bibinfo{author}{\bibfnamefont{R.~C.} \bibnamefont{Nichol}}, \bibnamefont{and} \bibinfo{author}{\bibfnamefont{P.}~\bibnamefont{V{\"a}is{\"a}nen}}, \bibinfo{journal}{Mon. Not. Roy. Astron. Soc.} \textbf{\bibinfo{volume}{467}}, \bibinfo{pages}{3239} (\bibinfo{year}{2017}), \eprint{1702.00418}.

\bibitem[{\citenamefont{Stern et~al.}(2010)\citenamefont{Stern, Jimenez, Verde, Kamionkowski, and Stanford}}]{Stern:2009ep}
\bibinfo{author}{\bibfnamefont{D.}~\bibnamefont{Stern}}, \bibinfo{author}{\bibfnamefont{R.}~\bibnamefont{Jimenez}}, \bibinfo{author}{\bibfnamefont{L.}~\bibnamefont{Verde}}, \bibinfo{author}{\bibfnamefont{M.}~\bibnamefont{Kamionkowski}}, \bibnamefont{and} \bibinfo{author}{\bibfnamefont{S.~A.} \bibnamefont{Stanford}}, \bibinfo{journal}{JCAP} \textbf{\bibinfo{volume}{02}}, \bibinfo{pages}{008} (\bibinfo{year}{2010}), \eprint{0907.3149}.

\bibitem[{\citenamefont{Borghi et~al.}(2022)\citenamefont{Borghi, Moresco, and Cimatti}}]{Borghi:2021rft}
\bibinfo{author}{\bibfnamefont{N.}~\bibnamefont{Borghi}}, \bibinfo{author}{\bibfnamefont{M.}~\bibnamefont{Moresco}}, \bibnamefont{and} \bibinfo{author}{\bibfnamefont{A.}~\bibnamefont{Cimatti}}, \bibinfo{journal}{Astrophys. J. Lett.} \textbf{\bibinfo{volume}{928}}, \bibinfo{pages}{L4} (\bibinfo{year}{2022}), \eprint{2110.04304}.

\bibitem[{\citenamefont{Moresco}(2015)}]{Moresco:2015cya}
\bibinfo{author}{\bibfnamefont{M.}~\bibnamefont{Moresco}}, \bibinfo{journal}{Mon. Not. Roy. Astron. Soc.} \textbf{\bibinfo{volume}{450}}, \bibinfo{pages}{L16} (\bibinfo{year}{2015}), \eprint{1503.01116}.

\bibitem[{\citenamefont{Kunz et~al.}(2006)\citenamefont{Kunz, Trotta, and Parkinson}}]{Kunz:2006mc}
\bibinfo{author}{\bibfnamefont{M.}~\bibnamefont{Kunz}}, \bibinfo{author}{\bibfnamefont{R.}~\bibnamefont{Trotta}}, \bibnamefont{and} \bibinfo{author}{\bibfnamefont{D.}~\bibnamefont{Parkinson}}, \bibinfo{journal}{Phys. Rev. D} \textbf{\bibinfo{volume}{74}}, \bibinfo{pages}{023503} (\bibinfo{year}{2006}), \eprint{astro-ph/0602378}.

\bibitem[{\citenamefont{Liddle}(2007)}]{Liddle:2007fy}
\bibinfo{author}{\bibfnamefont{A.~R.} \bibnamefont{Liddle}}, \bibinfo{journal}{Mon. Not. Roy. Astron. Soc.} \textbf{\bibinfo{volume}{377}}, \bibinfo{pages}{L74} (\bibinfo{year}{2007}), \eprint{astro-ph/0701113}.

\bibitem[{\citenamefont{Biesiada}(2007)}]{Biesiada:2007um}
\bibinfo{author}{\bibfnamefont{M.}~\bibnamefont{Biesiada}}, \bibinfo{journal}{JCAP} \textbf{\bibinfo{volume}{02}}, \bibinfo{pages}{003} (\bibinfo{year}{2007}), \eprint{astro-ph/0701721}.

\bibitem[{\citenamefont{Szydlowski et~al.}(2006)\citenamefont{Szydlowski, Stachowiak, and Wojtak}}]{Szydlowski:2005kv}
\bibinfo{author}{\bibfnamefont{M.}~\bibnamefont{Szydlowski}}, \bibinfo{author}{\bibfnamefont{T.}~\bibnamefont{Stachowiak}}, \bibnamefont{and} \bibinfo{author}{\bibfnamefont{R.}~\bibnamefont{Wojtak}}, \bibinfo{journal}{Phys. Rev. D} \textbf{\bibinfo{volume}{73}}, \bibinfo{pages}{063516} (\bibinfo{year}{2006}), \eprint{astro-ph/0511650}.

\bibitem[{\citenamefont{Szydlowski and Kurek}(2006)}]{Szydlowski:2006pz}
\bibinfo{author}{\bibfnamefont{M.}~\bibnamefont{Szydlowski}} \bibnamefont{and} \bibinfo{author}{\bibfnamefont{A.}~\bibnamefont{Kurek}}, \bibinfo{journal}{AIP Conf. Proc.} \textbf{\bibinfo{volume}{861}}, \bibinfo{pages}{1031} (\bibinfo{year}{2006}), \eprint{astro-ph/0603538}.

\bibitem[{\citenamefont{Burnham and Anderson}(2004)}]{Burnham:2002book}
\bibinfo{author}{\bibfnamefont{K.}~\bibnamefont{Burnham}} \bibnamefont{and} \bibinfo{author}{\bibfnamefont{D.}~\bibnamefont{Anderson}}, \bibinfo{journal}{A Practical Information-theoretic Approach}  (\bibinfo{year}{2004}).

\bibitem[{\citenamefont{Sugiura}(1978)}]{Sugiura01011978}
\bibinfo{author}{\bibfnamefont{N.}~\bibnamefont{Sugiura}}, \bibinfo{journal}{Communications in Statistics - Theory and Methods} \textbf{\bibinfo{volume}{7}}, \bibinfo{pages}{13} (\bibinfo{year}{1978}).

\bibitem[{\citenamefont{Di~Valentino et~al.}(2025)}]{CosmoVerseNetwork:2025alb}
\bibinfo{author}{\bibfnamefont{E.}~\bibnamefont{Di~Valentino}} \bibnamefont{et~al.} (\bibinfo{collaboration}{CosmoVerse Network}), \bibinfo{journal}{Phys. Dark Univ.} \textbf{\bibinfo{volume}{49}}, \bibinfo{pages}{101965} (\bibinfo{year}{2025}), \eprint{2504.01669}.

\bibitem[{\citenamefont{Rebou{\c{c}}as et~al.}(2025)\citenamefont{Rebou{\c{c}}as, de~Souza, Zhong, Miranda, and Rosenfeld}}]{Reboucas:2024smm}
\bibinfo{author}{\bibfnamefont{J.}~\bibnamefont{Rebou{\c{c}}as}}, \bibinfo{author}{\bibfnamefont{D.~H.~F.} \bibnamefont{de~Souza}}, \bibinfo{author}{\bibfnamefont{K.}~\bibnamefont{Zhong}}, \bibinfo{author}{\bibfnamefont{V.}~\bibnamefont{Miranda}}, \bibnamefont{and} \bibinfo{author}{\bibfnamefont{R.}~\bibnamefont{Rosenfeld}}, \bibinfo{journal}{JCAP} \textbf{\bibinfo{volume}{02}}, \bibinfo{pages}{024} (\bibinfo{year}{2025}), \eprint{2408.14628}.

\bibitem[{\citenamefont{Colg{\'a}in et~al.}(2025{\natexlab{b}})\citenamefont{Colg{\'a}in, Pourojaghi, and Sheikh-Jabbari}}]{Colgain:2025fct}
\bibinfo{author}{\bibfnamefont{E.~{\'O}.} \bibnamefont{Colg{\'a}in}}, \bibinfo{author}{\bibfnamefont{S.}~\bibnamefont{Pourojaghi}}, \bibnamefont{and} \bibinfo{author}{\bibfnamefont{M.~M.} \bibnamefont{Sheikh-Jabbari}} (\bibinfo{year}{2025}{\natexlab{b}}), \eprint{2505.19029}.

\end{thebibliography}
\end{document}